\documentclass[11pt]{article}
\usepackage{times}
\usepackage{graphics}
\usepackage{tensor}
\usepackage{natbib}
\usepackage{makeidx}
\usepackage{latexsym}
\usepackage{bm}
\usepackage{amsmath}
\usepackage{mathtools}
\usepackage{amsthm}
\usepackage{amssymb}
\usepackage{algpseudocode}
\usepackage{hyperref}
\makeindex

\newcommand{\beq}{\begin{equation}}
\newcommand{\eeq}{\end{equation}}
\newcommand{\dif}[2]{\frac{{\rm d} #1}{{\rm d} #2}}

\newcommand{\ildif}[2]{{\rm d} #1/{{\rm d} #2 }}

\newcommand{\ilpddif}[3]{\partial^2 #1/{\partial #2 \partial #3}}
\newcommand{\defn}{\begin{quote}{\bf Definition. }}
\newcommand{\edefn}{\end{quote}}
\newcommand{\thm}{\begin{theorem}}
\newcommand{\ethm}{\end{theorem}}

\newcommand{\bmat}[1]{\left ( \begin{array}{#1}}
\newcommand{\emat}{\end{array}\right )}
\newcommand{\E}{\mathbb{E}}

\newcommand{\ts}{^{\sf T}} 

\newcommand{\bp}{{\bm \beta}}

\theoremstyle{definition}

\theoremstyle{plain}
\newtheorem{theorem}{Theorem}

\newcommand{\eps}[3]
{{\begin{center}
 \rotatebox{#1}{\scalebox{#2}{\includegraphics{#3}}}
 \end{center}}
}

\begin{document}
\title{Some statistical aspects of the Covid-19 response
}
\author{Simon N. Wood$^1$, Ernst C. Wit$^2$, Paul M. McKeigue$^3$, Danshu Hu$^1$, Beth Flood$^1$,\\Lauren Corcoran$^1$ and Thea Abou Jawad$^1$,  \\
1. School of Mathematics, University of Edinburgh, U.K. \\
2. Institute of Computing, Univerit\`a della Svizzera italiana, Lugano, Switzerland\\
3. College of Medicine and Veterinary Medicine, University of
Edinburgh, U.K.\\
 {\tt simon.wood@ed.ac.uk}\\
}

\maketitle

\begin{abstract}
This paper discusses some statistical aspects of the UK Covid-19 pandemic response, focussing particularly on cases where we believe that a statistically questionable approach or presentation has had a substantial impact on public perception, or government policy, or both. We discuss the presentation of statistics relating to Covid risk, and the risk of the response measures, arguing that biases tended to operate in opposite directions, overplaying Covid risk and underplaying the response risks. We also discuss some issues around presentation of life loss data, excess deaths and the use of case data. The consequences of neglect of most individual variability from epidemic models, alongside the consequences of some other statistically important omissions are also covered. Finally the evidence for full stay at home lockdowns having been necessary to reverse waves of infection is examined, with new analyses provided for a number of European countries. \\ 

\smallskip
\noindent Read before The Royal Statistical Society at the Discussion Meeting on ‘Some statistical aspects of the Covid-19 response’ held at Hallam Conference Centre in London on Thursday, 10 April 2025, the President, Professor Sir John Aston, in the Chair. The paper attracted over 50 written comments. Our response to these is included after the main paper.
\end{abstract}

\section{Introduction}

Covid-19 caused immense strain on health systems and societies worldwide with the WHO official death toll to date\footnote{time of writing May 2023.} -- generally considered a lower bound -- corresponding to almost 0.1\% of the world population or a life loss of around 3 days per capita. Although governments around the world undertook strenuous measures to mitigate the threat, these did not come without side effects. The UK Covid response caused substantial collateral damage. Directly in terms of blocked or delayed access to healthcare \citep[e.g.][]{riera2021delays}, exacerbation of mental health problems \citep[e.g.][]{OConnor2021mental}, lost schooling and normal social development for children \citep[e.g.][]{major2021learning-loss}, and all the other human cost of cutting the great majority of the population off from normal social contact for months, under regulations that at times confined people to their own homes for 23 hours per day and prohibited outside contact other than online. Indirectly, and perhaps more substantially, through the human effects of the economic disruption, which the bank of England estimated was the largest for some 300 years, for the UK, and was unprecedented in that it involved deliberate halting of much economic activity, with money creation being employed to attempt to mitigate the consequent immediate problems. Creating money while reducing real economic  activity is obviously inflationary \citep[e.g.][Ch.1]{wolf23inflation,hall2009inflation} even without the supply chain problems \citep[e.g.][]{govt-supply-chain} that followed on the resumption of paused activity\footnote{to quote former BoE chair Mervyn King (BBC, 23/10/2022): `during Covid, when the economy was actually contracting because of lockdown, central banks decided it was a good time to print a lot of money \ldots That led to inflation. We had too much money chasing too few goods, and the result was inflation. That was predictable. It was predicted, and it happened.' }. The subsequent sharp increase in inflation\footnote{close to 7\% and increasing steeply before exacerbation by the Ukraine war.} is one path by which the disruption has contributed to increased economic deprivation \citep[e.g.][]{shine22inflation-poverty, richardson2023inflation-deaths} of the sort clearly linked to substantially reduced life expectancy and quality of life \citep[e.g.][]{marmot-review-10}. 

Some proportion of these effects would have happened under any response to Covid, and they obviously have to be considered against the reduction in human suffering that the Covid response achieved or might reasonably have achieved. In section \ref{sec:le} we argue that historic data on life expectancy reductions precipitated by large economic shocks, make it unclear whether the measures that were adopted will end up being net life savers, or achieve reasonably close to minimum achievable total life loss. Another indication of the reality of the trade-offs is that any reasonable estimate of the cost per life year saved from Covid by non-pharmaceutical interventions substantially exceeds the $\pounds 30\text{K}$ per life year threshold usually applied by NICE (the  UK National Institute for Health and Care Excellence) when approving introduction of a pharmaceutical intervention. For example, taking the half a million potential Covid deaths initially predicted for the UK under minimal mitigation \citep{rep9} less the recorded Covid deaths to date, would suggest around 300 thousand lives potentially saved. Given approximately one decade of life lost on average per victim \citep[e.g.][]{hanlon2020}, this corresponds to 3 million life years saved (16 life days per head). Taking this as the upper bound on life years saved, and a conservative $\pounds 10^{12}$ of extra borrowing plus lost economic activity as the cost of the interventions, gives a cost per life year saved over 10 times the NICE threshold. Given that health spending is necessarily finite, this comparison suggests a trade-off between life years saved from Covid versus life years saved from other diseases that may not be straightforward to justify.

If one accepts the existence of significant trade-offs, the reality of the collateral damage and its non-negligible size in relation to the benefits of the Covid measures, then in addition to identifying the many things that went well in the Covid response, it is important to discuss what went badly. Without such a dialogue we reduce the chance of doing better next time. This paper is about openly discussing statistical aspects of some of what did not go so well. Data and code used in the paper are supplied in a supplementary R package.

\section{The presentation of risk \label{risk}}

\begin{figure}
\eps{-90}{.55}{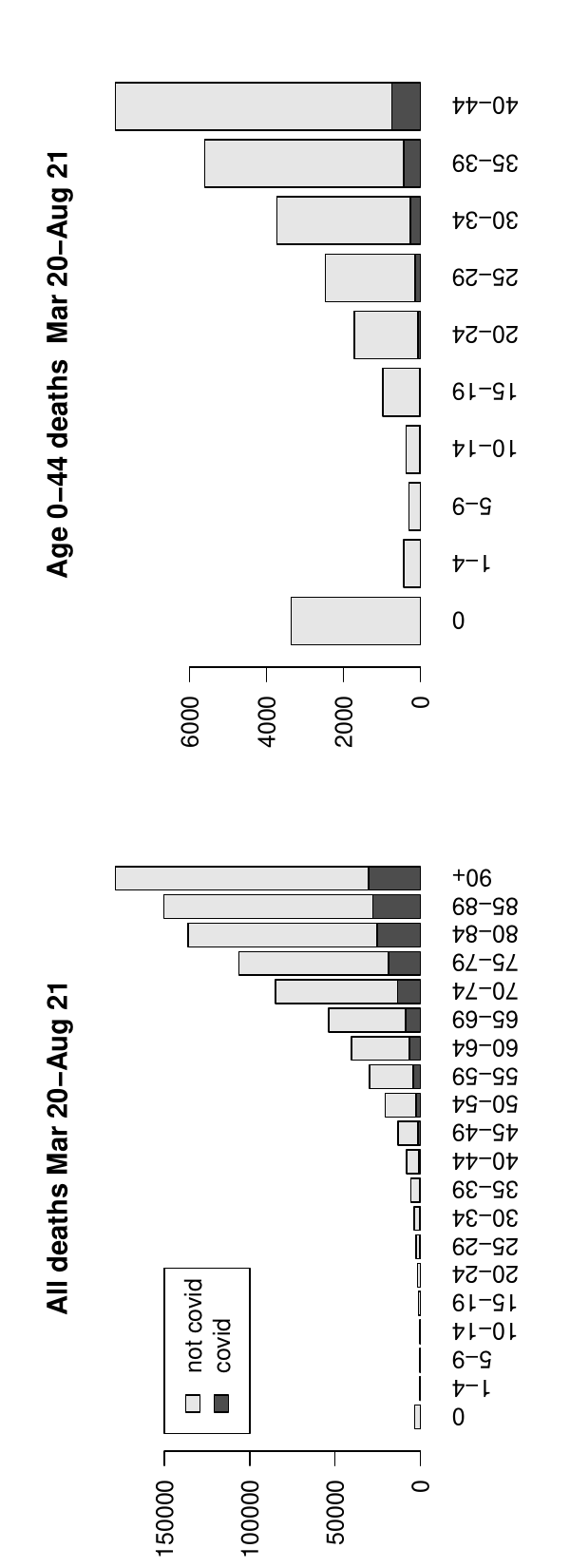}
\caption{All UK deaths from March 2020 to August 2021 by age band. Covid deaths are shown in black. Data are from the Office for National Statistics. The right panel simply enlarges the plot for the under 45s, where the risks are otherwise too low to be visible. \label{death-age}}
\end{figure}

\begin{quote}
\ldots a substantial number of people still do not feel sufficiently personally threatened; it could be that they are reassured by the low death rate in their demographic group\ldots the perceived level of personal threat needs to be increased among those who are complacent, using hard hitting emotional messaging.
\end{quote}
This unusual approach, of apparently intentionally distorting the presentation of medical risk in the service of a public health goal, is extracted from the 22 March 2020 recommendations of the UK government advisory Scientific Pandemic Influenza Group on Behaviour (SPI-B). The actual risk profile underlying the complacency is shown in figure \ref{death-age}. The ethics of distorting risk perception in this way are open to question, particularly if done in the interests of promoting a `greater good', subjected to detailed quantification of the short term benefits, but not of the long term disbenefits. One of the milder examples of the approach was a widely displayed government poster picturing a healthy woman in her mid twenties in a mask with the slogan `I wear this to protect you. Please wear yours to protect me.' The framing in terms of reciprocal risk implied either large overstatement of the risk to the person pictured, or a failure to consider her Covid risk in relation to her baseline risk. For example, the current best estimate for the return time of a super-volcanic eruption of the civilization ending magnitude that city dwellers are unlikely to survive is 17 thousand years \citep{rougier2018serisk}. Even only considering the two years of the pandemic this is likely larger than the Covid risk to the woman pictured. 

Figure \ref{death-age} is based on retrospective data, and one moderately frequent argument is that at the start of the UK epidemic the risk factors were not known. This is at best only partially true. Initially, the Chinese authorities were more open in sharing clinical data \citep[e.g.][online publication: Jan 30, Feb 7, Mar 9, Mar 9]{huang2020clinical,wang2020clinical, wu2020covid, zhou2020clinical} than they later were when it came to information pertaining to Covid origins. Statistical uncertainty notwithstanding, these studies provided a good deal of information on co-morbidity risk factors. Combined with the Diamond Princess cruise ship outbreak data, the studies also gave reasonably solid data on risk with age profiles \citep[e.g.][from March and May 2020]{verity2020ifr,lancet-ifr}.

The remainder of this section discusses how Covid risks were assessed and presented beyond the start of the first wave, as well as how authorities typically failed to present the health risks associated with the response.

\subsection{The risk from long Covid}

From mid 2020 the narrative around risk in the young and healthy began to shift towards the need to avoid long Covid, but again the presentation of the risk was questionable, and again hard hitting emotional messaging was used, rather than presenting the actual risk. For example, in October 2021 the Department of Health and Social Care released a film on the dangers of long Covid, the press release including this warning about the risks after mild illness:
\begin{quote}
Tom, 32, who features in the film\footnote{{https://www.gov.uk/government/news/health-secretary-warns-of-long-term-effects-of-covid-19-as-new-film-released and https://www.youtube.com/watch?v=ulJSEo2fWvA}. Tom is shown clearly face on and there is no indication that his name has been changed or that he  is being represented by an actor. He is also shown as having a Reading Half Marathon 2019 medal, but there are photos online of all the 4 Toms who completed that year, and none appear to be him (ditto Thomas or Tomasz).} says:
``Do not make the mistake of thinking that being young or being fit is going to stop COVID from having a long-term impact on your health.''
\end{quote}  

 As expected for any serious pneumonia \citep[e.g.][]{hopkins1999ards,herridge2011pneumonia}, hospitalised patients clearly suffered long term effects \citep[][for example\footnote{although with a general population control group this does not provide evidence for a Covid specific syndrome.}]{ayoubkhani2021post}, and a novel virus was always likely to result in an increase in people suffering longer term post-viral complications \citep[e.g.][]{appelman2024long-covid}. However, beyond anosmia (loss or change in sense of smell) the evidence for exceptional risk from Covid specific sequelae after mild illness came from studies with substantial statistical problems, and it is difficult to view the level of concern about such sequelae as reflecting normal evidence based medicine. One problem is the tendency to use catch all definitions such as the \cite{NICE-long-covid} definition of {\em Post COVID-19 Syndrome:}
\begin{quote}
Signs and symptoms that develop during or after an infection consistent with COVID‑19, continue for more than 12 weeks and are
not explained by an alternative diagnosis.
\end{quote}
Obviously such a definition invites {\em post hoc ergo propter hoc}.  Almost any event could be substituted for {\em infection consistent with COVID‑19} and a substantial number of cases of the associated syndrome would be found. High prevalence estimates were almost always based on a wide range of self reported symptoms, often in non-representative samples, with no control group, so that a meta-analysis of 174 studies produced up to Janaury 2022 \citep{omahoney-lc-meta} reported at least 45\% of Covid cases producing ongoing symptoms after 4 months.
\cite{amin2021long-covid}, \cite{haslam2023long-covid} and \cite{hoeg2023long-covid} all highlight the lack of control groups in most of these early studies, the vague and variable case definitions, and the use of samples that were not representative of the general population. As an illustration of the problems, an ONS study \citep{ONS-long-covid} found prevalence of long Covid symptoms at 5\% in confirmed Covid cases versus 3.4\% in matched controls, suggesting a 95\% CI for Covid associated symptom prevalence of (1.0, 2.2)\%. The same study found {\em self reported} long Covid prevalence of 11.7\%. Nevertheless, the ONS continued to publish a survey of self reported long Covid until March 2023 when a prevalence of about 2.9\% was reported \citep{ONS-long-covid-survey}.

A further difficulty, even with carefully designed studies, is relatively low response rates and the associated serious risk of participation bias. For example of the approximately 800,000 REACT study participants invited to take part in the REACT Long Covid study 276,840 responded \citep{react-long-covid}: there is an obvious danger that those with ongoing symptoms are more likely to participate. Similarly a large scale Scottish study \citep{hastie-scot-lc} had response rates of 15\% for the controls and 20\% for the cases: it only takes a proportion of the extra 5\% in the case group to have participated {\em because} of symptoms, to produce the differential symptom rates observed between cases and controls, even if the real rates are identical. Prospective studies should in principle be more reliable. The prospective cohort case control study of \cite{ballering-lc} found 381/1782 Covid cases with at least one ongoing symptom at 90-150 days post infection, versus 361/4130 for a control cohort matched by `event' time, binarized age and sex, suggesting a Covid attributable rate of 12.7\%, in a 6:4 female skewed cohort with age distribution over-concentrated around mean 54. But drop out was high. The initial Covid cohort size was 4,231, so that there is considerable scope for differential drop out between those with and without ongoing post Covid symptoms to have skewed the rates\footnote{e.g. if the symptom rate was 9\% in cases and controls, but almost all those with ongoing symptoms post Covid stayed in then the results would be very close to those obtained. Loss of interest drop out in the fully recovered is also not implausible.}. Perhaps more concerning is that, except for ageusia/anosmia, the `core symptoms' employed in the paper are all from the somatization sub-scale of the Symptom Checklist-90 questionnaire, used to assess psychological problems which may be expressed somatically \citep[see e.g.][]{holi1998soma, vandriel2018soma}. To be sure that these are physical sequelae to Covid itself requires subject blinding to their Covid status. In any case, the initial cohort included 142 hospitalized cases (final number not given) and the final cohort contained 158 cases with ageusia/anosmia. Hence, from the information presented, it is not possible to work out rates for Covid attributable symptoms excluding ageusia/anosmia in non-hospitalized patients, or even to rule out these being zero.

In fact, for non-hospitalized cases, carefully designed studies often found low, or even no, difference in the frequency of persistent symptoms in cases and controls, except for anosmia. For example, a systematic review and meta-analysis of 22 studies of children and young people \citep{behnood2022long-covid} reported that the frequency of most reported persistent symptoms was similar in SARS-CoV-2 positive cases and controls. A study in Norway \citep{selvakumar2023long-covid} using the WHO definition of `post-COVID-19 condition' found that prevalence at 6 months was similar in test-positive non-hospitalized cases and test-negative controls. In a large population-based French cohort for ages 18-69 \citep{matta2022long-covid}, self-reported Covid-19 infection was associated with persistence of multiple physical symptoms 10-12 months after the first wave, whereas laboratory-confirmed Covid-19 infection was associated only with anosmia.  The authors suggest that persistent symptoms `may be associated more with the belief in having experienced COVID-19 infection than with actually being infected with the SARS-CoV-2 virus'. In a study comparing ongoing symptoms post Covid and post Influenza, \cite{brown2023-covid-flu} found no difference in symptom rates between the two cohorts.  None of this is to deny the existence of real sequelae to Covid infection, but rather to suggest that the long Covid evidence base was of insufficient statistical quality to form part of the justification for continuation of severe societal restrictions, or of risk distorting public messaging .  Looking forward, \cite{hoeg2023long-covid} make recommendations for improving the situation, to both avoid irresponsible exaggeration of risk after mild infection, and to better serve those suffering long term problems genuinely caused or triggered by Covid infection.

The referees suggested also discussing the economic effect of long Covid, in particular on UK employment levels. The background is that people of working age not in employment as a result of long term illness has increased by around 700,000 since the start of the pandemic, an increase of around a third of the pre-pandemic level. \cite{ONS-inactivity-health} and the associated data set reports estimated increases in health related inactivity by primary cause, stating that long Covid is classified under `Other health problems or disabilities' which increased by 156,000 between 2019 and 2023. This is some 80,000 more than would be expected from the general increase, suggesting an upper bound for long Covid's impact on long term inactivity. 
An alternative estimate comes from \cite{ONS-long-covid-labour} which estimated a 0.5-3.4\% increase in economic inactivity among those self reporting long Covid suggesting 10-70 thousand people with long Covid related inactivity if we take the contemporaneous ONS estimate of 2 million people with self reported long Covid as all of working age (an overestimate). For comparison, 593 thousand report musculoskeletal or connective tissue problems and 645 thousand depression or other mental disorders as their primary causes of economic inactivity. Of course those economically inactive from long Covid represent only a proportion of those substantially impacted, but this is equally true of the other medical causes of inactivity.


\subsection{The risk of life loss from economic shock\label{sec:le}}

\begin{figure}
\eps{0}{.55}{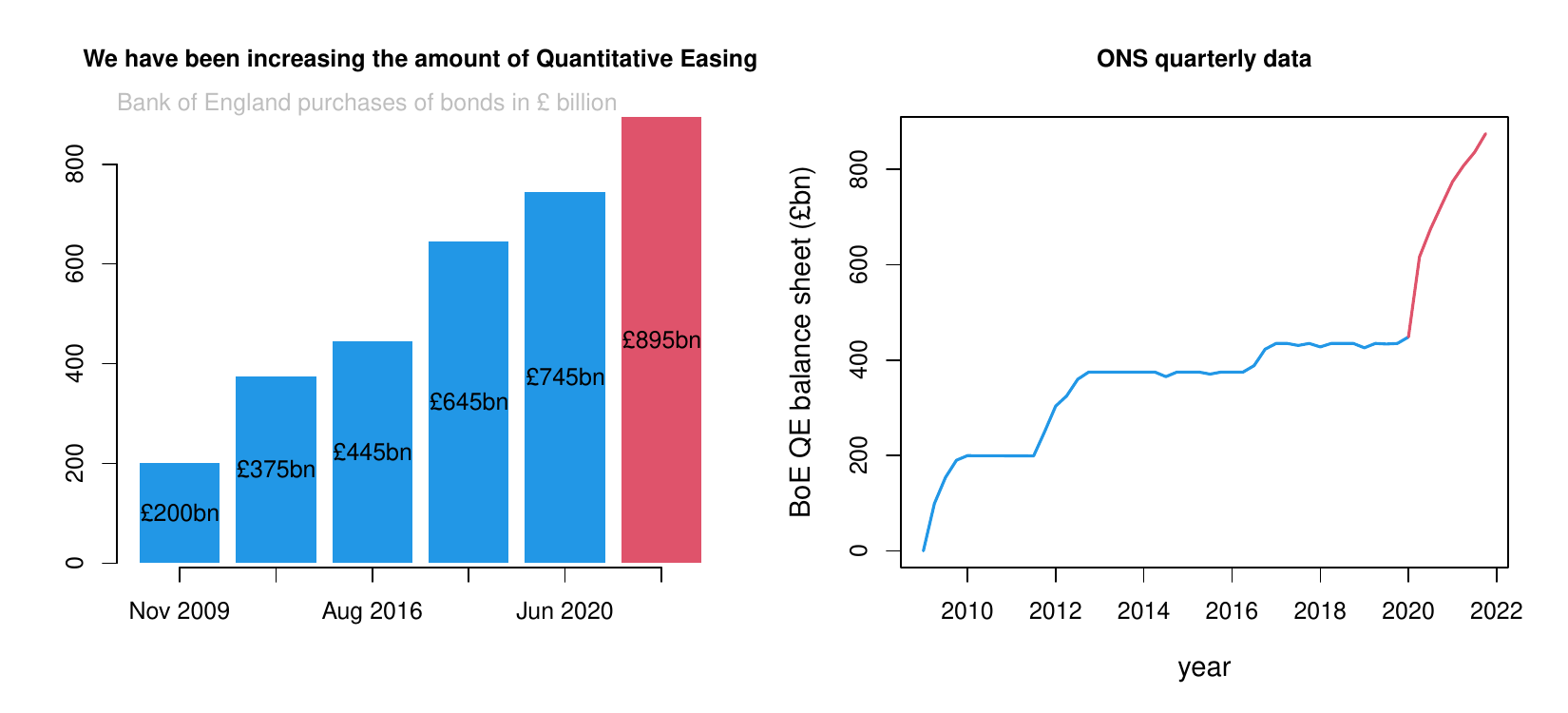}
\vspace*{-.5cm}

\caption{Left: The Bank of England quantitative easing (QE) program as presented on the Bank of England website in 2021 (redrawn, but identical in all relevant respects). Right: the same QE program using a continuous linear time scale (ONS data), with the Covid part shown in red. \label{boe-qe}}
\end{figure}

While {\em some} of the risks from Covid were being exaggerated, the risks from the measures taken were downplayed. The left panel of figure \ref{boe-qe} is faithfully redrawn from the Bank of England website\footnote{{\tt www.bankofengland.co.uk/monetary-policy/quantitative-easing} -- graphic removed late 2022.}. Its rescaling of the time axis creates the impression that the quantitative easing program (funded by money creation) had expanded steadily over time, with the Covid part simply following an existing trend. The right panel, using a continuous linear time axis to plot the same data, creates a rather different impression.

Discussions that involved the size of the QE program were often framed in terms of `saving life versus saving the economy'. Such framing relies on not considering another set of data: that relating to the health effects of economic deprivation and inequality. A major evidence based review of these links had been published early in 2020, before the pandemic \citep{marmot-review-10}. That review updated an equally substantial 2010 report on the links between economic deprivation and health inequality. The 2020 report includes a forensic, data-based investigation of how health outcomes and life expectancy for the more disadvantaged were worsened by the exacerbation of economic deprivation following the financial crisis of 2008 and subsequent government response to it (the latter democratically sanctioned by the 2010 election\footnote{see, e.g. p. viii 2010 Conservative party manifesto, p14 2010 Liberal Democrat Manifesto}). 

\begin{figure}
\eps{0}{.6}{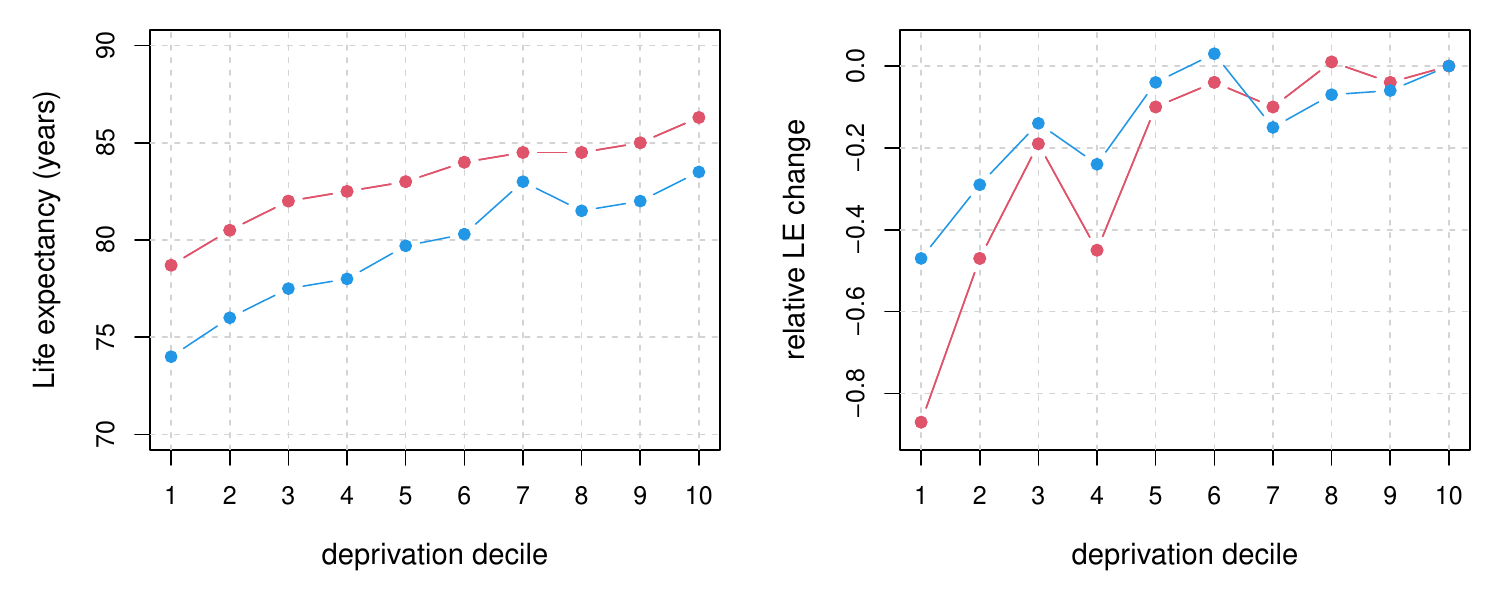}
\vspace*{-.5cm}

\caption{Left: Life expectancy by (population weighted) area deprivation decile for women (red) and men (blue). Right the change in years of life expectancy relative to the least deprived between 2010-12  and 2016-18 for women (red) and men (blue). Data from figures 2.3 and 2.5 of \cite{marmot-review-10}, which were based on ONS and PHE data.
 \label{marmot}}
\end{figure}

The left panel of figure \ref{marmot} shows how life expectancy changes with population weighted decile of area deprivation in England \citep[from][Fig 2.3]{marmot-review-10}. Taking the upper 10\% as representing current potentially achievable life span, the plotted data correspond to a, presumably avoidable, life loss of 3.2 years per capita for women and 3.95 years for men. If the mean life expectancy of the upper half is taken as the achievable figure, then the loss reduces to about 1.8 and 2.5 years respectively. Scaling up to the current UK population, this is a potentially avoidable life loss of 140-290 million years. Whether one is politically inclined to view this life loss as avoidable or not, any small percentage increase to it, as a result of the economic disruption caused by the Covid response, obviously risks life loss on a scale comparable to that from Covid itself. 

How likely is such an increase in life loss? Internationally there are clear examples of linkage between economic problems, deprivation and lifespan reduction \citep{ruminska1997econ-death, ciment1999life, case2017mortality}.  Also, \cite{Stuckler2010crises} point out an association between welfare spending and mortality strong enough, if it contains a causal component, to imply that any crisis either directly reducing spending or preventing increased spending would result in substantial avoidable life loss\footnote{They found a 1.2\% reduction in all cause mortality for each 100USD per capita increase in welfare spending. The increase of around \pounds 1000 per head in annual government debt servicing costs since 2020 is obviously unavailable for increased welfare.}. Recent historical data for the UK also give some indication. \cite{marmot-review-10} argue that the response to the 2008 financial crisis is implicated in a reduction in the trend for increasing life expectancy equivalent to around 1 year life loss per capita. If the larger financial shock from the Covid response produces knock on effects anything close to this, then the measures will have cost far more life than they saved. However, this figure is based on assuming that an apparent linear trend in life expectancy before 2010 would simply have continued afterwards, in the absence of the 2008 crisis and response. What actually happened to life expectancy is being compared to the extrapolation of a purely statistical model. 

A much more conservative approach is to base all comparisons on what actually happened to life expectancies: to take the post-2008 life expectancy trends among the more prosperous as the measure of what could reasonably be expected for those not substantially impacted by economic effects, rather than straight line extrapolations. Then we can ask how the economic deprivation life expectancy gap in fact changed after the 2008 crisis. 
Treating the least deprived 10\% as the `more prosperous' group, the right hand panel of figure \ref{marmot} shows the results of such a difference in differences approach. The corresponding increase in avoidable life loss is about 7-9 weeks per capita, or 9-12 million life years for the whole UK population, and the evidence presented in \cite{marmot-review-10} makes it difficult to discount the knock on effects of the crisis having had a very direct role in this. It is unclear that there are any solid reasons to expect the Covid response economic deprivation related life loss effects to be smaller. 

One objection to this difference in differences approach is that for the most part life expectancy improved somewhat across the deprivation scale after 2008, but more slowly for the more deprived. The argument is therefore that the loss is not `real'. But such a position has perverse consequences. Even post-2008, average UK life expectancies grew by around 18 days per year. Is any per capita life loss in a year that is less than this not real? The direct loss from Covid over 2020 amounted to about 6 days per capita, and it would be absurd to say that there was no real loss of life because this is less than the expected 18 day life expectancy increase for the year. Similarly, deprivation has caused real loss of life, even though the average life expectancy has increased. 

A final objection is that the exacerbation of economic deprivation after a financial shock is not inevitable, but is a political choice: hence these data are irrelevant in considering whether to impose the shock, and are only relevant to how its consequences are later dealt with. This seems to us to be a counsel of perfection, firstly because there are inevitably some economic constraints on what is politically possible (money spent on debt servicing is not available for welfare, for example), and secondly because the historic data probably offer a better guide to what is politically likely than utopian considerations of what political action could ideally achieve. Whatever one's views about the policies adopted in the UK after 2008, they are what the electorate explicitly voted for in 2010. In other words, there is a real risk that the electorate may not vote for the `best' policies after an economic shock, or that no political party offers such policies. Should such a real world risk be ignored? 

This is of course not to argue that a substantial risk equates to certainty. After the profound existential trauma and economic dislocation of the second world war the electorate did vote for a substantial enhancement of the welfare state. However this expansion cost less than 3\% of GDP when annual growth was 3-4\%, and substantial US aid was available, so that, despite the economic situation, government debt began falling sharply immediately after the war ended \citep[see][for more detail]{crafts2023welfare}. The contrast with current circumstances implies a need for circumspection in reading this precedent. The postwar reforms also point, perhaps, to the limits of what even very enlightened policy can achieve, post shock. Although the detailed mechanisms are of course complex, the ratio of standardized mortality rates in social classes V to I had actually increased in 1949-53 relative to the pre-war figure \citep{pamuk1985social}. Similarly the government commissioned 1980 Black report\footnote{{https://sochealth.co.uk/national-health-service/public-health-and-wellbeing/poverty-and-inequality/the-black-report-1980}} found that health inequalities had increased since 1948 \citep{gray1982onblack}. The fact that Black's recommendations for improvement were not implemented, while many of the later \cite{acheson1998} report were, may at least partly reflect the difference between what was politically possible in the recession of the early 1980s versus the strong economy of 1998, although differences in government (and electorate) political philosophy were also significant. 

The economic deprivation figures and the very approximate 3 million life years potentially losable to Covid were available by mid March 2020. Clearly the path from economic shock to substantial life loss is uncertain and very difficult to credibly model. This does not mean that there is no evidence that the effect exists. It only means that it comes with a range of possible loss of life and associated uncertainty, which is the very basis of the definition of risk. In short, the available data in early 2020 indicated that a large economic shock would come with substantial risk of downstream loss of life.

\subsection{The meaning of life expectancy}

The life expectancies discussed above broadly have the interpretation that if things stay much as they have been over the preceding few years, then this is how long we can expect to live, {\em provided nothing changes drastically}. The changes in life-expectancy discussed by \cite{marmot-review-10} are of this slowly varying nature, which is why it is reasonable to use them in calculations involving potentially avoidable life year loss. However, during the pandemic, media reports often stated that Covid had caused a life expectancy drop of around one year, while almost always omitting the qualifier that to interpret this as indicating that the average UK resident's expected lifespan had been shortened by a year would be entirely false. The one year drop is what would happen if there was a new Covid epidemic, causing comparable life loss, every year from 2020 onwards. Unqualified statements such as
\begin{quote}
 {Americans are now expected to live an average of 77.3 years, down from 78.8 years in 2019}\footnote{https://www.cnbc.com/2021/07/21/life-expectancy-in-the-us-declined-in-2020-especially-among-people-of-color-.html}
 \end{quote}  or later and hence less dramatically, 
 \begin{quote}
 A boy born between 2018 and 2020 is expected to live until he is 79, down from 79.2 for the period of 2015-17\footnote{https://www.theguardian.com/society/2021/sep/27/covid-has-wiped-out-years-of-progress-on-life-expectancy-finds-study},
  \end{quote}
  not only omit the caveat but contradict it, explicitly interpreting {\em period} as {\em cohort} life expectancy. 

The media reports were based on scientific papers that perhaps assumed that their readership understood the caveat, without the need to explicitly state it. A case in point is \cite{islam2021effects}, who reported both reductions in `life expectancy' and life years lost for 37 countries for 2020. For example, they report a 2 year life expectancy drop for Bulgarian men, who were also estimated to have lost 7260 life years to Covid per 100 thousand population. The latter figure corresponds to an average life loss of 4 weeks per head. The equivalent figures for the UK in 2020 was a life expectancy drop of about 1 year and a life loss of about 6 days per head\footnote{The current total from the start of the pandemic stands at about 12 days per head, under the assumption of a decade of life loss per victim and given the UK government figure of 226 thousand deaths with Covid. }. When assessing risk, the difference between risks that would shorten your expected lifespan by 1 year versus 6 days is quite substantial. Presenting a figure in a way likely to lead to a 60 fold over-estimation of actual risk seems unlikely to promote a proportionate response to the risk.

\subsection{Excess deaths \label{sec:excess}}

\begin{figure}
\eps{0}{.6}{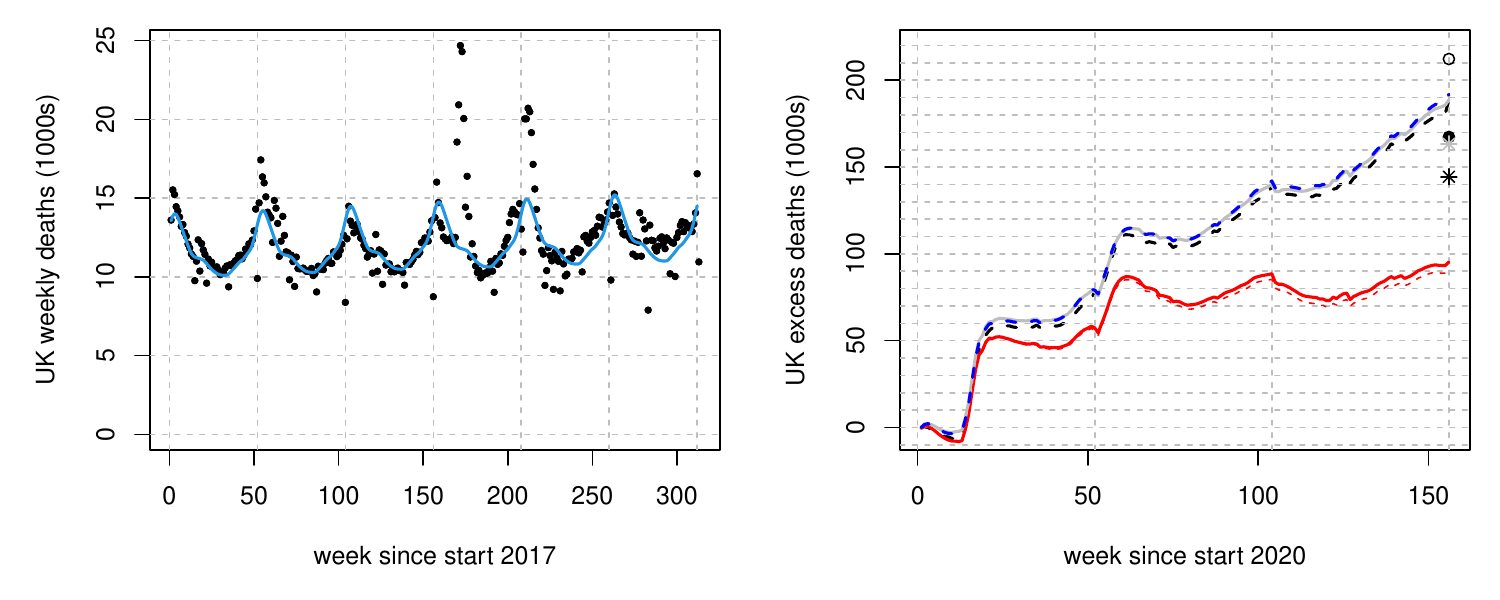}
\vspace*{-1cm}

\caption{Left: Weekly UK deaths (black dots), and weekly deaths predicted using 2017-19 life tables (blue). The predictions start from the age structure of the population at the start of 2017 for the first 3 years, and the age structure at the start of 2020, thereafter. The cumulative observed and predicted deaths match to within $0.0025\%$ (50 deaths) over 2017-19. Right: red solid is the cumulative excess deaths from the start of 2020 iterating from population by age data for the start of 2020; red dashed is the equivalent iterating 6 years from population by age data for the start of 2017; black dashed is for the conventional method based on 2017-19 data; grey is the same but using life table iterated weekly deaths in place of raw deaths; blue dash is the life table based approach with ageing turned off, starting from the estimated population by age in mid 2018. The open circle is the government figure for deaths with Covid, the black disc is the ONS pandemic excess death figure, the star is the PHE pandemic excess death figure for England (grey scaled up pro-rata to the UK). We argue that the red curve is the most reasonable, because it directly accounts for the effects of population ageing on the expected deaths. 
 \label{excess}}
\end{figure}

\begin{figure}
\eps{0}{.6}{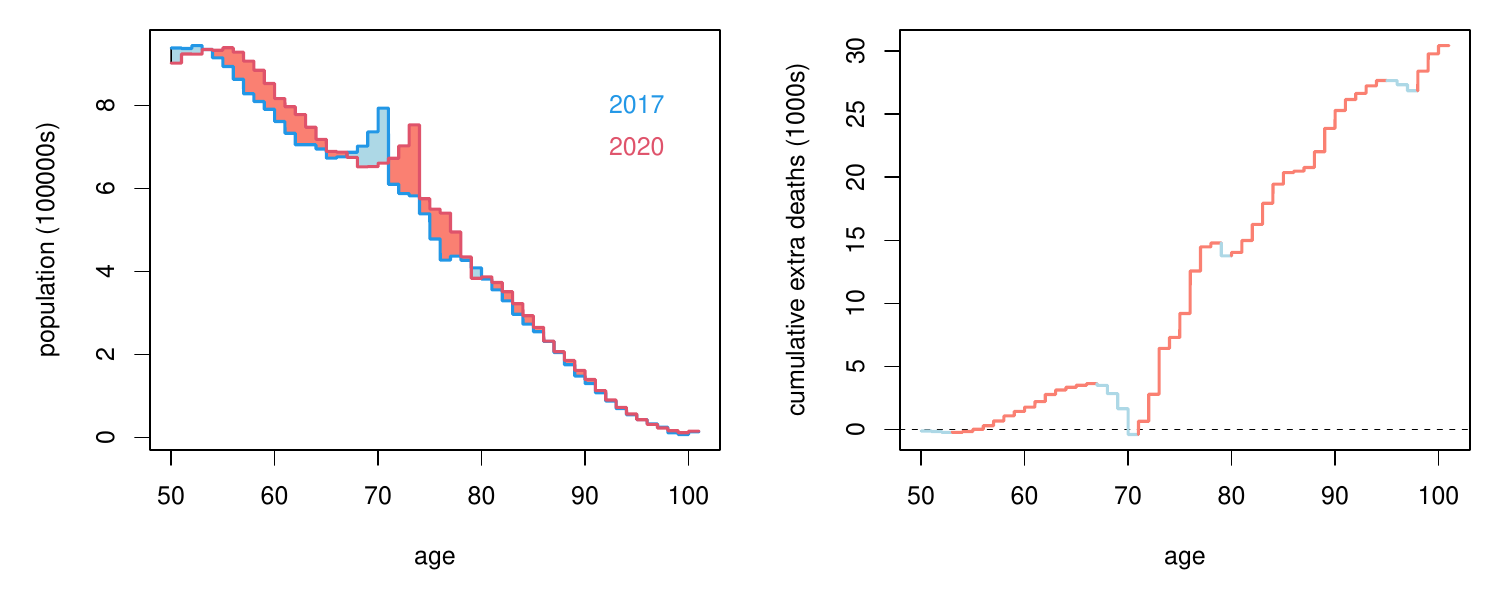}
\vspace*{-1cm}

\caption{The effects of population ageing on deaths. Left: blue shows the UK population by one year age band in 2017 and red shows the equivalent in 2020. The difference between the two ($\Delta_i$ for age class $i$) is shaded blue when the 2020 population is less than the 2017 population and salmon when the 2020 population is larger. Right: the population difference, $\Delta_i$, in each year class between 2020 and 2017 is multiplied by the annual probability of death in that class, $m_i$. The cumulative sum of the resulting expected extra deaths per year (i.e. the expected extra deaths among those from 50 up to age $i$, $d_{\le i} = \sum_{j=50}^i m_j \Delta_j$) is then plotted against age class, with colour coding corresponding to the left plot. The three years of population ageing leads to 30 thousand extra expected deaths, a large effect that can obviously not be captured by the traditional approaches to computing the expected deaths for excess death calculations.  Population data and annual per capita mortality rates at age are from the ONS.
 \label{ageing}}
\end{figure}

A useful way of calibrating Covid risk retrospectively is provided by computing excess deaths: the excess of actual deaths over what might reasonably be expected given the situation in previous years. But here too there is scope for statistical confusion. Excess deaths are often computed by looking at deaths relative to seasonal averages over a number of years preceding the period of interest, with various adjustments made to account for trends in mortality over time. That adjustments are needed is clear if the post-war baby boomers are considered. For the UK, the year group conceived immediately after the war is 31\% larger than the year group from the previous year (see figure \ref{ageing}). These people were approaching 75 at the start of 2020. Failure to take account of this demographic cliff edge advancing into the age group at which mortality rises sharply with age is bound to lead to inflation of apparent excess death rates. 

We argue that the simplest approach to excess deaths is to take life tables computed from the mortality data over a reference period of the years  immediately preceding the period of interest, along with the population's age structure at the start of the period of interest, and to simply iterate the ageing and death processes forward in time. Applying the same process from the population structure at the start of the reference period offers the sanity check that the total deaths over the reference period should match between data and predictions. See Appendix \ref{demog} for details.

Figure \ref{excess} does this for the UK, with 2017-19 as the reference period, and all data obtained from the Office for National Statistics. 
Ageing and death are applied weekly to weekly age cohorts. The yearly cycle is obtained by fitting a generalized additive model (GAM) to the reference period data with a smooth for week, and  a cyclic smooth for week of the year. From this, a multiplier relating the weekly death rate to the annual death rate can be computed, so that weekly death rates can be used in the iteration of the demography. Note that the total number of deaths predicted by this iteration over 2017-2019 matches the observed to within 50 deaths (0.0026\%), when the iteration is started from the age structured population at the start of 2017. The figure of around 95 thousand total excess deaths from the solid red curve in the right panel of figure \ref{excess} is lower than the figure of 167,356 given by the ONS as the excess death figure from March 2020 until the end of 2022, or the PHE figure for England of 144,446 (equivalent to around 163 thousand if crudely scaled up to the UK). The ONS figure is based on simply comparing weekly deaths to the average for that week of the year for the five years preceding 2020, with PHE similar but based on statistically modelled deaths. These figures obviously neglect the consequences of baby boomer and general population ageing that lifetable iteration includes. That this ageing effect is indeed large can be confirmed by a very simple check, which is shown graphically in figure \ref{ageing}. The left hand plot compares the population in each age class from age 50 onwards for 2017 and 2020 according to ONS. For each age group the growth in population from 2017 to 2020 can be multiplied by the annual death rate for that age group (from ONS life tables), and then summed to get the expected change in number of deaths per year that ageing has caused. The total is about 30 thousand expected extra deaths per year and the right hand plot, of cumulative changes in death with age, illustrates how the change is accumulated across the age groups. Note that migration makes negligible contribution to these figures given that only some 7\% of migrants are over 50 and `negligible' numbers over 70 \citep{migrants2020}.

Since the ONS and PHE figures are based on a 5 year time window, the right panel of figure \ref{excess} also shows the results of applying the standard weekly death rate averaging method using 2017-19 as the reference period. The total excess deaths predicted by this method actually exceed the ONS and PHE figures, which is unsurprising given that 2015 was a relatively high death year and 2019 relatively low. To further emphasise the dependence of the standard method on neglecting ageing, it is also possible to modify the iterated lifetable approach by turning off ageing (while also having deaths not deplete the age groups, so that the overall population does not decline). Applying this, obviously deficient, process, starting from the mid 2018 age structure, gives a close match to the current standard methods.

Given the size of the ageing effects, it is difficult to see that the iterated lifetable approach does not give the more reasonable expected number of deaths, relative to the current standard methods. The interpretation of the resulting expected deaths is also particularly clear: {\em if age specific death rates remained unchanged from the reference period then this is the number of deaths that would occur}\footnote{Note that, at about 140 per 100 thousand our figures are still higher than the 120 per 100 thousand estimated excess deaths over the two pandemic A/H3N2 influenza seasons starting in 1968 \citep{viboud2005flu68}}.  

The cumulative excess deaths shown in red in the right panel of figure \ref{excess} are much lower than the total deaths recorded with Covid (212,247 with Covid mentioned on the death certificate by the end of 2022, according to the UK government's data dashboard). There are a number of mechanisms that are likely to account for this. An obvious one is the fact that only some 17 thousand people had only Covid and nothing else recorded on their death certificate. When Covid is only one factor among several in a death, it is statistically naive to expect it to contribute a whole extra death in the excess figures (given that some of the other factors are risks contributing to what is expected without Covid). Put slightly more technically, since dying with Covid and dying with other co-morbidities are not independent events, Covid mortality events do not simply add to the mortality caused by the other co-morbidities. Related to this are what epidemiologists refer to as `harvesting' effects: where an epidemic pathogen brings forward the deaths of some very frail people by only a few weeks or months. Over a period of three years many such people will not appear as excess deaths at all, since their death has only been moved {\em within} the time period considered.

For near real time monitoring of excess deaths it has been argued \citep[e.g.][]{holleyman2022} that expected deaths should be corrected for harvesting effects - that someone who was expected to die shortly from other causes, but succumbs to Covid earlier, should then be removed from the later expected deaths. The approach obviously has some philosophical difficulties, since dying of {\em any} cause means that you do not die later of some other cause: hence a decision has to be made to treat only some causes of death in this way. We in effect {\em define} Covid deaths as excess\footnote{one problem with this is that many lives of the terminally frail are eventually ended by opportunistic infections: focusing on the particular infectious agent is not always meaningful in such cases.}. A second problem with using the approach to compute excess deaths over an extended period is the change in interpretation of the statistic. Conventionally excess deaths in a period is the excess of observed deaths, $D$, above the number expected over the period given the mortality data up to the start of the period, $E$. Consider the case in which some horrible event led to 10,000 unexpected deaths of otherwise healthy 5 year olds in a year. Both the Holleyman method and the conventional method count these as 10,000 excess deaths in the year. Now consider the case in which $D$ is 10,000 less than $E$, but 20,000 people die one day earlier than expected from the cause selected under the Holleyman method. The Holleyman method again produces 10,000 excess deaths, as under the first scenario, although in total we saw 10,000 fewer deaths than were expected to occur over the year. A further problem is what to do about the situation in which occurrences such as a low respiratory pathogen season (or lockdown) displaces some deaths to {\em later} than they would have been otherwise expected to occur. If we correct for people who die earlier than expected due to an unusually high risk from pathogens, should we not also correct for those who die later than expected due to an unusually low risk from (other) pathogens? 
Focusing on excess life year loss would avoid {\em some} of these problems, but is not easy, given available data, without strong modelling assumptions. 

An objection to the explanation that in many cases Covid may have brought forward deaths by `only' a few months, is that studies looking at life loss per Covid victim suggest figures of around a decade on average \citep[e.g.][]{hanlon2020}, which would imply a more limited role for harvesting. However such life loss studies tend to suffer from the problem of having to treat co-morbidities as simple categories (often binary), with limited possibility for incorporating co-morbidity severity, leading to likely inflation of the estimated life loss per victim. For example, suppose that each victim with congestive heart disease as a co-morbidity is assigned the average life expectancy of someone with congestive heart disease. Then we will over-estimate life loss to Covid, if in reality it is those with more severe disease, and consequently shorter Covid free life expectancy, who are most likely to succumb to Covid.   



\section{Covid cases and other media distortions}

\begin{figure}
\eps{0}{.54}{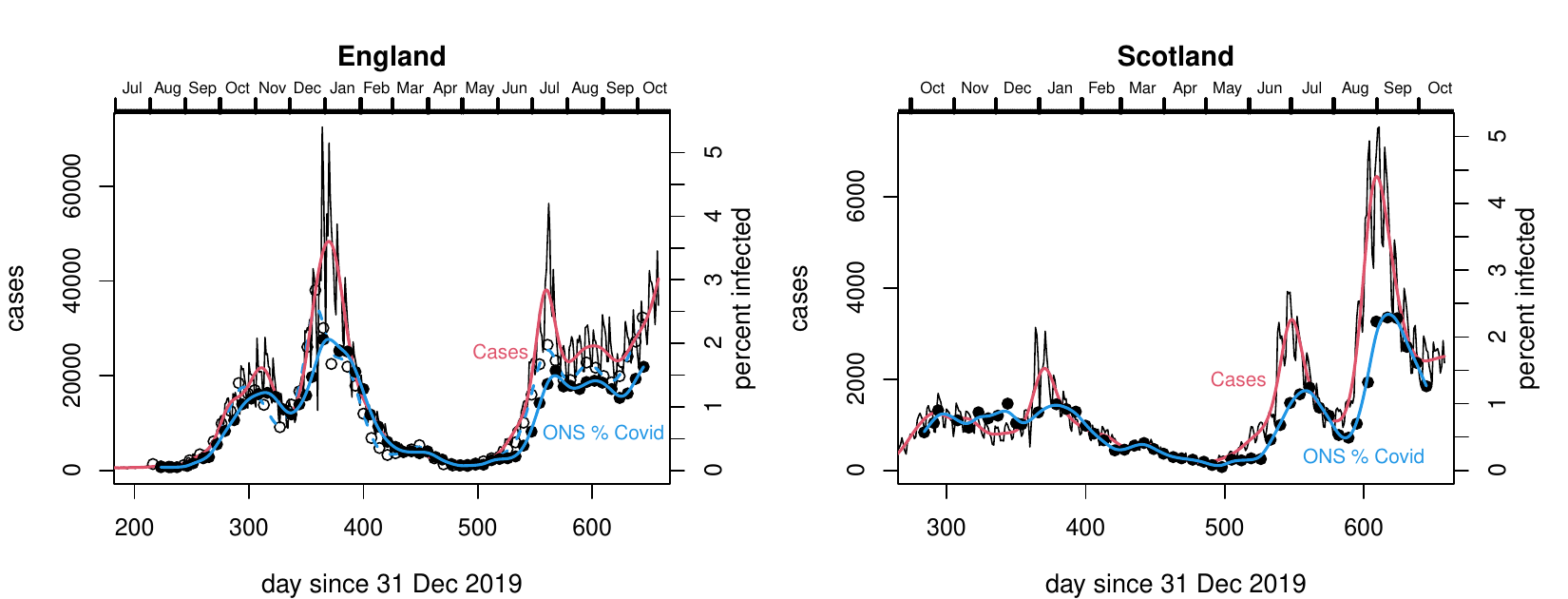}

\caption{Comparison of official daily Covid cases data (black lines, red smooth) and ONS prevalence measurements (black dots, blue smooth) for England on the left and Scotland on the right. In both cases the data are scaled to match over the 50 days from the midpoint day 430. If case data reliably measured prevalence then the smooth curves should be uniformly close to each other. For England the ONS incidence estimates (scaled) are also shown, as open circles smoothed by a dashed blue curve.   
 \label{cases-ons}}
\end{figure}
A statistically troubling feature of the media and government presentation of the state of the Covid pandemic was the preference given to `case' data, even after the ONS had started directly measuring prevalence using sampling, from mid 2020 onwards. Case data were discussed as if they were proportional to prevalence, although it is unclear what population they sample and how they relate to prevalence. The data measure people testing positive among those who were tested: largely those who chose to be tested, or were advised to by test and trace, and could obtain a test. How this number relates to prevalence at any given time is somewhat obscure: some relationship is to be expected, but it seems unlikely to be constant, in large part because of the way testing behaviour was likely to change over time, for example in relation to both perceived risk and available testing capacity. Figure \ref{cases-ons} shows the correspondence between prevalence measurements and case data over time, scaling the data to match over a time interval in the middle of the period. The focus on case data creates a clear danger of over-reacting when prevalence is increasing, as numbers of infections appear to climb more steeply than is actually the case. 

It can also be argued that case data should be proportional to incidence, not prevalence (although, for a disease of 2 weeks or so duration, the distinction is a rather fine one, given the difficulty identifying the population being sampled).  This introduces the additional problem that there is an epidemiologically significant and variable delay from infection (the event relevant for incidence) to detection as a case. The left panel of figure \ref{cases-ons} also shows the ONS surveillance survey reconstructions of incidence for England, again scaled to match cases over 50 days from the data-period  midpoint. Despite the generally poor correspondence, there are periods on the upswing of a wave when the cases match scaled incidence quite well, before the cases overshoot. This match probably results from the over acceleration of case detection compensating for the delay from infection to case detection - it is unlikely that anyone would argue that it is prudent to rely on such a fortuitous cancellation of biases. 

\begin{figure}
\eps{0}{.66}{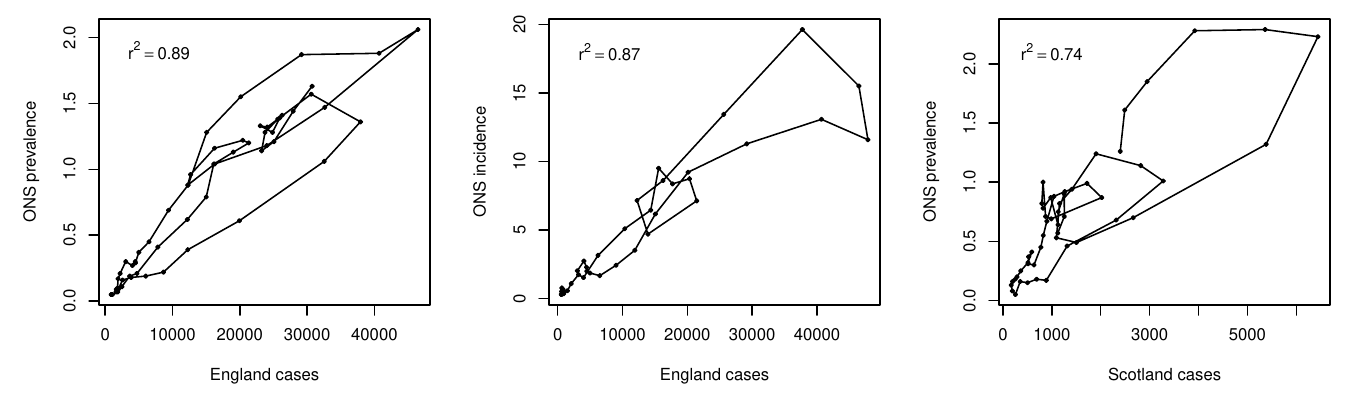}

\caption{Left: ONS prevalence plotted against cases (weekly cycle smoothed out) for England, with points joined in time order. Middle: similar, with ONS incidence on the vertical axis. Right: as left but for Scotland. The shifting nature of the relationship between cases and prevalence or incidence is clear.  
 \label{phase}}
\end{figure}

These problems have not prevented some authors from continuing to argue that cases represent a good proxy for actual prevalence or incidence. For example \cite{brainard2023cases-vs-ons} make this case on the basis of the correlation coefficients between cases and ONS estimates. Figure \ref{phase} illustrates the substantial systematic drifts in calibration that lie behind the correlations.  

A curious argument was sometimes advanced that the case data were reliable for assessing trends in infection rates and the pathogen reproductive number, $R$, {\em if looked at over a short enough time window}. This seems equivalent to falsely asserting that $\lim_{\Delta \to 0} \Delta^{-1}\{f(t+\Delta)+g(t+\Delta) - f(t)-g(t)\} = \ildif{f}{t}$, where $g(t)$ is some function deemed inconvenient. That concern over such matters is more than nit-picking is illustrated, for example, by \cite{liu2021NPIimpact}, who attempted to model the effects of various non-pharmaceutical interventions (NPIs) on $R$, which was estimated from case data. Their table 5 summarizes the effects of the 13 NPIs considered. All apparently reduced $R$, apart from test and trace, which apparently increased it. It seems improbable that test and trace actually increased transmission, and substantially more likely that it accelerated the finding of cases leading to inevitable upward bias in the case based $R$ estimate.

Poorly founded opinions are to social media what sand is to a beach, so we will not comment on what appeared there. But it is reasonable to expect better from traditional media with a reputation for journalistic integrity, fact checking and editorial control. This expectation was not met by, for example, the BBC's failure to put Covid deaths in the context of the average number of daily deaths from all causes, or by the Guardian having Nassim Taleb write about a historically rather moderate pandemic as if it were a `black swan event' -- a perspective unlikely to promote a balanced view of the actual risks. Another Guardian article from 27 April 2020\footnote{https://www.theguardian.com/commentisfree/2020/apr/27/business-lockdown-johnson-tory-donors} compared Covid to the black death, concluding that Covid was in some ways worse and
\begin{quote}
\ldots if we ``open up the economy'' to help Tory grandees make money, there won't be much economy left once the second wave of infections has finally settled down, because all the Topshops will have to be razed to make space for graves.
\end{quote} 
The black death is estimated to have killed 30-60\% of Europe's population irrespective of age. For Covid in the UK to be somehow equivalent it would have had to cause a life loss of some 20 years per capita. The current figure actually stands at less than 2 weeks per capita. Even at the end of 2021 exaggeration was still  common: for example the {\em yahoo!news} headline `Omicron: Germany records highest COVID daily death toll in nine months' was difficult to view as balanced at a point in time at which Germany had 4 Omicron cases, all very much alive. Most media commentators of course avoided such statistically nonsensical hyperbole, but the more moderate often still repeated that Covid was `the worst pandemic for a century', an odd view at a time when LBGT rights and black lives matter were to the fore in public consciousness: the WHO estimates the AIDS death toll at 27-48 million, again with a high life year loss burden per death. 

In some ways a more statistically concerning example, was a 19 April 2020\footnote{https://www.theguardian.com/commentisfree/2020/apr/19/coronavirus-deaths-data-uk} article in the Guardian from two professors of biostatistics discussing the supposed difficulty of estimating prevalence:
\begin{quote}
Arguably, the most important problem is the ``denominator'' - what is the actual number of people who are infected by the virus? This is virtually impossible to determine, except perhaps in the unlikely scenario of real-time, continuous, population-wide testing.
\end{quote} 
The Guardian was not interested in printing a short letter correcting this statistically unusual perspective, by pointing out that randomized sampling could be used (as the ONS and REACT subsequently did).

That the media gets things wrong is perhaps why we have independent fact checkers, but it is not clear that they too were not over-hasty in their judgements about which Covid narratives were `correct', at least in part reflecting a lack of statistical knowledge. For example, in a lengthy article on evidence for lockdown efficacy {\tt fullfact.org} asserted that the reason \cite{wood2020arxiv} had UK new infections per day peaking well before lockdown, while \cite{flaxman2020lockdown} had surging growth in infections until lockdown (see section \ref{lockdown}) was that the former had assumed a much longer infection to death duration than the latter. That would be a compelling argument, were it not for the fact that the papers had used essentially the same infection to death distribution \citep{verity2020ifr}. {\tt fullfact.org} were not interested in correcting this, nor some other incorrect statements about infection to death timings relating to apparently not understanding right truncation. They were also not interested in updating their article in the light of the REACT-2 and ONS incidence reconstructions covered in section \ref{lockdown}. 

This type of misleading and selective use of statistical evidence was not limited to the media. For example in 2021 the official online Scottish government advice on face coverings stated that 
\begin{quote}
Scientific evidence and clinical and public health advice is clear that face coverings are an important part of stopping the spread of coronavirus.
\end{quote}
and provided a link for the scientific evidence. This turned out to be a SPI-B/SAGE advice summary\footnote{{\tt https://assets.publishing.service.gov.uk/government/uploads/system/uploads/
attachment\_data/file/948607/s0995-mitigations-to-reduce-transmission-of-the-new-variant.pdf}}, which cited two pieces of scientific evidence, apparently suggesting transmission reductions from mask wearing of 6-15\%, or up to 45\%, respectively. The paper cited as evidence for the first figure was in fact an editorial \citep{cowling2020masks}, which also pointed out that the paper cited for the 45\% figure \citep{mitze2020masks} was flawed (the design appears unable to pick up the case in which mask wearing is actually harmful, for example). The editorial's figure is quoting a properly conducted meta-analysis \citep{brainard2020masks}  which actually concluded
\begin{quote}
\ldots wearing a mask may slightly reduce the odds of primary infection with [Influenza Like Illness] by around 6 to 15\% [\ldots] This was low-quality evidence\ldots
\end{quote}

\section{Epidemic dynamic models \label{models}}

Perhaps the most surprising feature of the epidemic models used to justify Covid policy was the omission of the fundamental role of person-to-person transmission rate heterogeneity investigated by \cite{novozhilov2008}, and explicitly raised as a serious issue for Covid models in early 2020 by Gomes \citep[eventually published as][]{gomes2022}. The degree of variability between people in their susceptibility, connectivity and other determinants of transmission probability profoundly affects the size of epidemic -- or of epidemic waves -- predicted by the Susceptible Exposed Infectious Recovered (SEIR - exposed are infected but not yet infectious; recovered actually includes dead) type models that were used. The mechanism is simple: those individuals most susceptible to infection or most socially connected are preferentially removed from the susceptible population first, leading to a much more rapid reduction in infection rates than simple depletion of a susceptible population of clones would produce. Realistic levels of variability can easily halve the predicted epidemic (wave) size, and yet the models only accounted for the very modest heterogeneity in mean contact rates with age. It is possible that the early work on this topic was inaccessible, so we present the mathematical fundamentals of the mechanism here.

\subsection{Person-to-person variability in SEIR models}

First, let $\alpha$ be a parameter determining susceptibility to infection, which varies over the susceptible population, and let $s(\alpha,t)$ denote the susceptible population per unit $\alpha $ interval with parameter $\alpha$ at time $t$. Without loss of generality we can scale the problem so that the initial population is 1, in which case $s(\alpha,0)$ is the initial p.d.f. of $\alpha$. The standard SEIR model for this situation is,
$$
\dif{s(\alpha,t)}{t} = - \alpha s(\alpha,t)I(t), ~~~ \dif{e(\alpha,t)}{t} =\alpha s(\alpha,t)I(t) - \delta e(\alpha,t),~~~
\dif{i(\alpha,t)}{t} = \delta e(\alpha,t) - \gamma i(\alpha,t)
$$   
where $I(t) = \int i(\alpha,t) d\alpha$. On integrating the first ODE we have $s(\alpha,t) = s(\alpha,0) \exp( - \alpha q_t)$ where $q_t = \int_0^t I(t^\prime) dt^\prime$. Since $q_t$ is monotonic in $t$, it is immediately clear how the epidemic progresses faster in subpopulations with higher $\alpha$. {\em In itself this observation suggests that great care is needed in extrapolating to the whole population from those who become sick first.} 

We can now obtain the total susceptible population at time $t$ by integrating out $\alpha$ 
$$
S_t = \int s(\alpha,0) \exp (- \alpha q_t) d \alpha = M(-q_t)
$$
where $M$ is the moment generating function of the initial distribution of $\alpha$ (by definition). Now consider the time derivative of $S_t$,
$$
\dif{S}{t} = - \int \alpha s(\alpha,t) d \alpha I(t) = -M^\prime(-q_t) I(t) = -M^\prime \{ M^{-1}(S_t)\} I(t)
$$ 
where $M^{-1}$ is the inverse function of $M$. So the SEIR dynamics are determined by three ODEs, without explicit dependence on $\alpha$.

Variability in contact rates can be modelled in a similar way. It is assumed that transmission depends on the product of $\alpha$ for the susceptible and $\alpha^\prime$ for the infected. In this case
$$
\dif{s(\alpha,t)}{t} = - \int \alpha \alpha^\prime s(\alpha,t)i(\alpha^\prime,t) d \alpha^\prime = -\alpha \bar \alpha^\prime_t s(\alpha,t) I(t)
$$
where $\bar \alpha^\prime_t = \int \alpha^\prime i(\alpha^\prime,t)/I(t)d \alpha^\prime $. Analytical progress now requires the approximation that the infectious state is short enough that the distribution of $\alpha$ in the infectious stage at $t$ is proportional to the distribution in those first becoming infected at $t$. That is $i(\alpha,t) = \alpha s(\alpha,t)$, so that $\bar \alpha^\prime_t = \int \alpha^2 s(\alpha,t) d\alpha / \int \alpha s(\alpha,t) d \alpha $. If we now redefine $q_t = \int_0^t \bar \alpha_{t^\prime}^\prime I(t^\prime) d \alpha^\prime$, then the maths follows through similarly to the variable susceptibility case, with the addition that $\bar \alpha^\prime_t = M^{\prime\prime}(-q_t)/M^\prime(-q_t)$, so that we end up with 
$$
\dif{S}{t} = -M^{\prime\prime} \{ M^{-1}(S_t)\} I(t),
$$ 
and again the original infinite dimensional system is reduced to three ODEs.

\begin{figure}
\eps{-90}{.7}{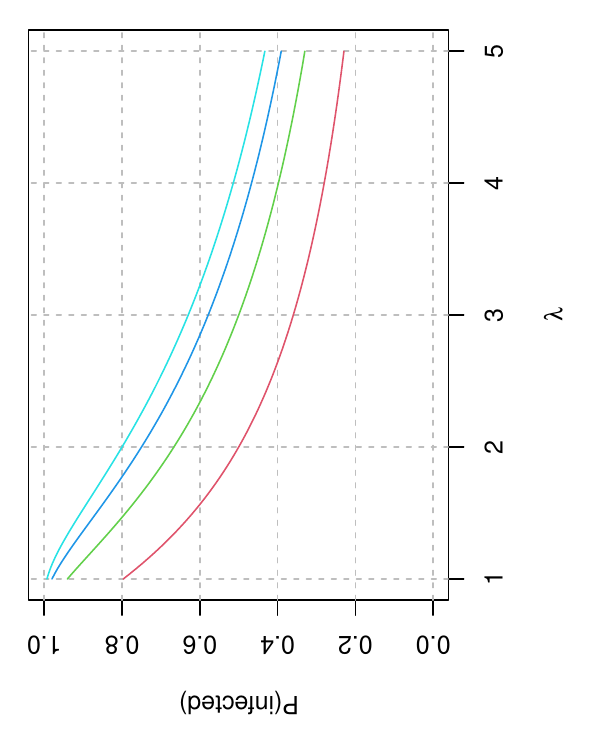}
\vspace*{-.6cm}
\caption{Final proportion infected against  $\lambda$ for $R_0=2,3,4,5$ (ascending, red, green, blue, turquoise), for SEIR epidemic models with susceptibility or mixing rates varying across individuals. The age band dependent variability used by SAGE models in the UK corresponds to $\lambda \approx 1.2$. Realistic estimates seem to be in the range 2.5-5. \label{episize} }
\end{figure}

If $\alpha$ has a $\text{gamma}(k,\nu)$ distribution  in the initial susceptible population, with p.d.f. $\nu^k \alpha^{k-1}e^{-\alpha\nu}/\Gamma(k)$, then under either model we have 
$$
\dif{S}{t} = - {R}_0 \gamma S^\lambda_t I(t)
$$
where $R_0$ is the initial pathogen reproductive number and the `immunity coefficient' $\lambda = 1+1/k$, or $1+2/k$, for the variable susceptibility or variable contact rate models, respectively. \cite{novozhilov2008} demonstrates that this is also a good approximation for a wide variety of other initial $\alpha$ distributions. Integration and routine re-arrangement then shows that the final proportion infected, $x$, must satisfy
$$
x = 1-\{1+(\lambda-1)R_0 x\}^{-1/(\lambda-1)}.
$$
Figure \ref{episize} plots the final proportion infected against  $\lambda$ for several values of $R_0$. \cite{gomes2022} estimate that $\lambda=2.9$ for England and Scotland, and that the heterogeneity with age generally assumed in SAGE modelling corresponds $\lambda \approx 1.2$. \cite{tkachenko2021} estimates $\lambda $ between 4.1 and 4.7 for several US cities. In the case of transient immunity subsequent waves are to be expected, of course, but the basic mechanism applies to each of them. 

It is possible that this effect was neglected because it was felt to be difficult or impossible to estimate the person to person  variability in transmission rates, but neglecting an effect known to have a large unidirectional effect on results simply adds spurious precision to estimates that are then almost bound to be wrong. In any case, not including this effect in models calibrated against data is puzzling. 

\subsection{Other major modelling omissions \label{models2}}

Another oddity of models used to try to infer $R$, or detect the effect of lockdowns from data, was the fact that they did not model the rather fundamental division of the population into locked-down and key-worker compartments. Those compartments must have different transmission rates, with the difference increasing with the efficacy of lockdown as a suppression measure. The decision was presumably made on grounds of simplicity and the lack of data sufficiently disaggregated to be informative about rate differences between the compartments. In itself this simplification may be a reasonable judgement call, but it does imply the need for care to ensure that the model formulation retains the flexibility to deal with the consequences of the simplification. 

One such consequence concerns the modelling of $R$ after lockdown. $R$ measures the average number of new infections caused by each existing infection. Crucially it is the {\em population of infections} that is being averaged over, not the population of people. Immediately after lockdown $R$ is depressed in the locked-down population, increasingly so with time as the infectious run out of household members to infect. Meanwhile the key-worker population maintains a higher $R$. Since initially most infections are in the locked-down compartment the whole population $R$ initially tracks what is happening in that compartment, but over time an ever increasing proportion of infections is in the key-worker population, so that over time $R$ drifts upwards towards the key-worker compartment $R$. To avoid artefacts, models without separate locked down and key worker compartments clearly need to include an $R$ model flexible enough to capture this expected dip and recovery in $R$. \cite{Birrell2021} did this, but the highly cited \cite{flaxman2020lockdown} did not, instead assuming that $R$ was constant during lockdown: the serious artefacts that this induced are discussed in section \ref{alt-ld}.

Nosocomial infection (disease transmission in hospitals) was also absent from the models despite \cite{wang2020clinical} reporting a suspected 41\% nosocomial infection rate in Wuhan as a key finding in early February 2020, a feature that would be repeated in the first wave in Lombardy in Italy where \cite{boccia2020italy} note that  ``SARS-CoV-2 became largely a nosocomial infection''.  Later analysis showed that within Scotland the proportion of serious Covid that was hospital acquired peaked at around 60\% \citep{mckeigue2021shield}. Model based analyses, whether statistical or not, are likely to be severely compromised if such a significant transmission route is omitted.

All that said, almost certainly the most important omissions were the negative collateral impacts of the interventions. Of course there are good reasons why these effects were not included in the epidemic models themselves. But failure to  put as much effort into assessing the negative side effects of interventions as was devoted to predicting positive Covid reduction effects, is likely to have biased decision making in a manner unlikely to have achieved anything close to minimum practical societal loss. For example, the July 2020 Government report that attempted some quantification of negative impacts of lockdown \citep{covid-life-loss-govt} did not attempt any quantification of effects beyond 5 years.  For shorter timescales a much more speculative approach (appendix D4 of the report) was employed than for the disease modelling. This produced results  very difficult to reconcile with what actually happened post 2008, as recorded in the data discussed in section \ref{sec:le}.
\section{Lockdowns \label{lockdown}}

\begin{figure}
\eps{-90}{.65}{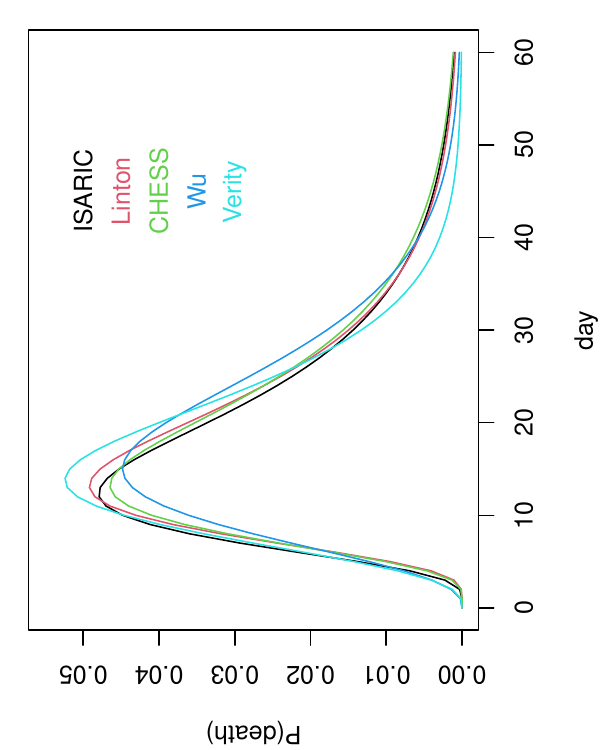}
\caption{Comparison of onset to death duration distributions from various sources. The ISARIC distribution is used here, because it is based on by far the largest sample size. Note that the distributions are not significantly different given the sample sizes involved.\label{isaric}}
\end{figure}

That the extreme reduction in contact rates accompanying lockdowns would suppress transmission rates and likely lead infections to decline rather than increase is uncontroversial. However the retrospective  belief that lockdowns were {\em necessary} for infection levels to fall is a conclusion that seems to be based on informal reasoning and a priori modelling, rather than data. The informal reasoning is approximately as follows
\begin{quote}
Across a large number of countries the same pattern was always seen: cases and deaths were increasing until the government imposed a full stay at home lockdown. Only then did cases and deaths decline. Clearly lockdowns caused the decrease, where all preceding measures had failed.  
\end{quote}
This argument is flawed. Full lockdowns are drastic measures of last resort. As such no government would impose them unless cases and/or deaths were still increasing, and they were necessarily the last measure imposed. But cases and deaths had to decline eventually. In consequence the pattern of increase-lockdown-decrease is simply inevitable and conveys no information about lockdowns' role in reversing waves of infection, no matter how often the pattern is repeated. The view that nothing preceding lockdown had worked, because cases and deaths were still increasing until lockdown, neglects the fact that cases and deaths are lagged data, only occurring around one or more weeks after infection. It is what the {\em current} daily new infection (incidence) rate is doing that indicates the success or otherwise of {\em current} measures. {\em Current} cases and deaths can not do this.

\begin{figure}
\eps{-90}{.545}{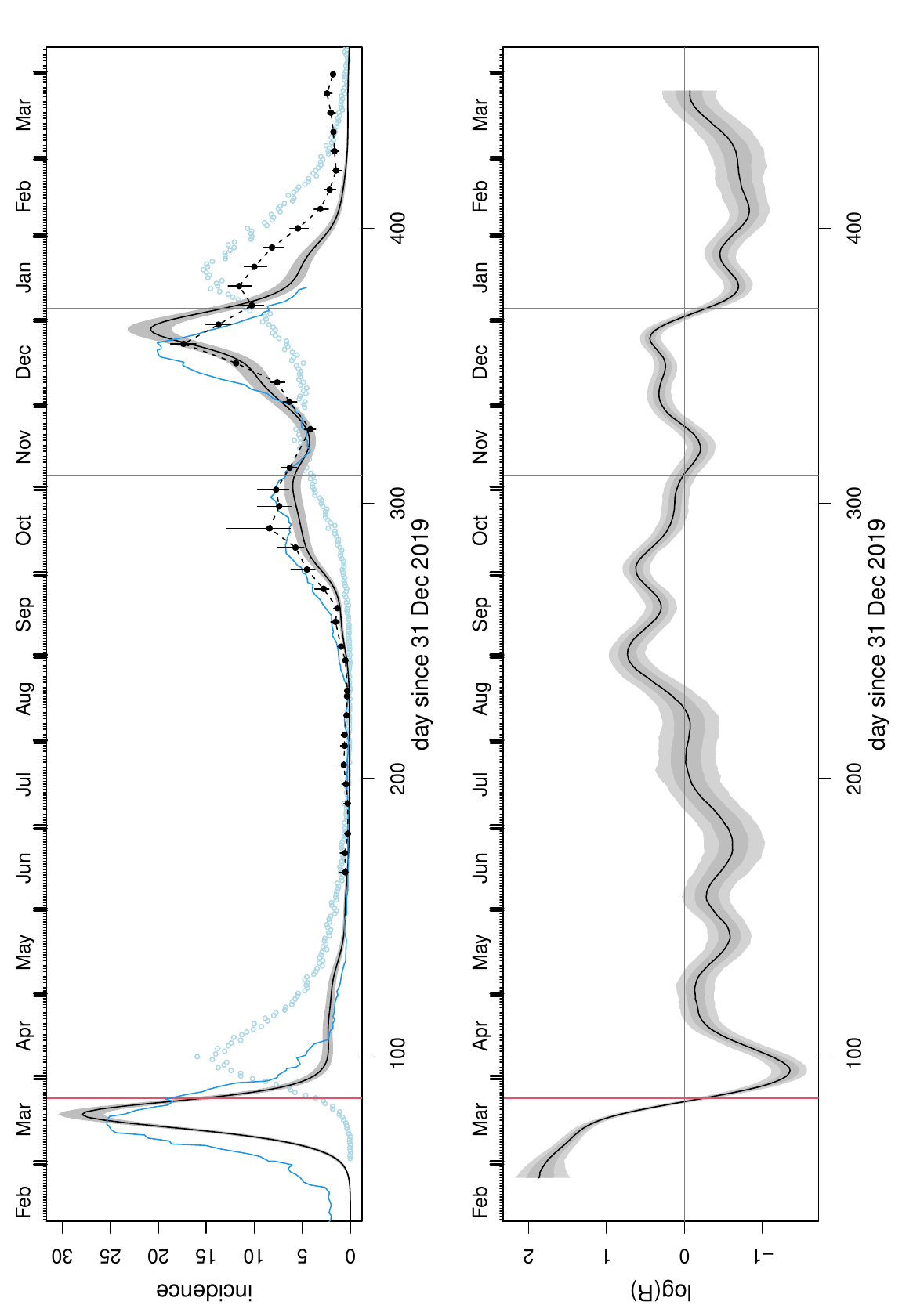}
\caption{Top: The grey bands are 95\% credible intervals for fatal incidence (new infections per day) per million, reconstructed from the NHS England daily hospital death data shown as light blue circles. The dark blue curve shows (scaled) reconstructed incidence from the REACT-2 study's random sample of the English population (the study reconstructed incidence of first symptoms, which has been lagged by the 5.8 day average delay from infection to first symptoms). The black dots with confidence bars, joined by dashed lines are the ONS reconstructions of incidence (scaled) from their randomized surveillance sampling data. Bottom: Natural log of the pathogen reproductive number, $R$, obtained from the grey incidence curve. \label{ir-eng}}
\end{figure}

Incidence is difficult to observe directly, but it is possible to retrospectively infer incidence trajectories consistent with the observed daily deaths from Covid. Criteria, ascertainment fraction and distribution of time from infection to event are clearly understood and relatively constant for deaths (none of these things would be true of cases). The simplest approach uses a basic deconvolution model \citep{wood2020arxiv,wood2021peakTiming}. Suppose that $y_i$ is the number of Covid deaths occurring on day $t_i$, then 
$$
\E(y_i) = \sum_{d=0}^{D_i} \exp \{f(t_i-d)\} \pi(d),
$$ 
$f(t)$ is the smooth log fatal incidence rate at day $t$ and $\pi(d)$ is the probability of an infection to death time interval of $d$ days. $D_i$ is the maximum lag from infection to death considered. To promote statistical stability, at the start of the epidemic this may be set to somewhere around 20 days, since the first deaths observed will tend to be from shorter duration disease. $D_i$ then grows at a day per day up to some limit (e.g. 80 days). This approach avoids estimating $f$ over a long initial period where $\exp(f)$ is essentially zero. The model can also be multiplied by a second log cyclic smooth term, to deal with the slight weekly cycle in deaths seen in some countries. $y_i$ can be assumed to follow a negative binomial or Poisson distribution. The smooth terms in the model can be represented using cubic splines with smoothing parameters estimated by (Laplace approximate) REML. Assuming smoothness on the log scale mitigates against the possibility of smoothing artefacts driven by rapid changes in absolute incidence. Appendix \ref{smoothing} provides more detail.

The infection to death distribution is available from several sources. The meta-analysis of \cite{mcaloon20} combines studies to provide an estimate of the distribution of time from infection to symptom onset. \cite{verity2020ifr}, \cite{linton2020incubation} and \cite{wu2020covid} all provide estimates of the distribution of time from symptom onset to death while properly accounting for right truncation in the data used. Given relatively small sample sizes in these early studies, \cite{wood2021peakTiming} integrated the uncertainty in the distributions into the analysis. The results were also compared to those obtained to fitting a model to CHESS data, although it was not possible to obtain data with nosocomial infections filtered out, so that a mixture model approach was necessary. However, later \cite{isaric.oct20} provided results on time from hospitalization to death, and from symptom onset to hospitalization, for a sample of  24,421 fatal cases across multiple countries (the sample is dominated by wealthy countries), at a point in time at which right truncation was a minor issue. Incorporating the \cite{mcaloon20} results, the corresponding infection to death duration model is $\log(d) \sim N(3.151,0.469^2)$, and given the large underlying sample size, is the model used here. Figure \ref{isaric} compares the various onset to death distributions. 

\subsection{Fatal incidence in England}

The reconstructed fatal incidence curve for England is shown in the upper panel of figure \ref{ir-eng}, based on NHS England hospital deaths data, with the corresponding log of the pathogen reproductive number $R$ shown below. $R$ can be obtained from the incidence curve by assuming a simple SEIR model as described in \cite{wood2021peakTiming}. The results for the first lockdown are very similar to those obtained by the beginning of May 2020. Later, two more direct statistical reconstructions of incidence became available. The most direct came from the REACT-2 study \citep{ward2021react}. Subjects in the study's random sample of English residents who tested positive for SARS-CoV-2 antibodies were asked when their symptoms started. This provides an estimate of the number of newly symptomatic cases each day, from which incidence can be obtained, by applying the same deconvolution method used with the deaths, or simply by lagging the curve by the mean infection to onset duration \citep[5.8 days according to][]{mcaloon20}. The blue curve in the upper panel of figure \ref{ir-eng} shows the result, digitized from \cite{ward2021react}. Note the somewhat high estimates very early on -- presumably representing misattribution of symptoms from other respiratory ailments to Covid, chiefly among subjects whose Covid infection was asymptomatic or very mild. However, even if these `background infections' were completely suppressed by pre-lockdown behaviour, there are not enough of them to alone account for the overall pre-lockdown drop, without Covid infections also having been in decline. The Office for National Statistics also published incidence reconstructions based on their large scale randomized surveillance sample from June 2020 (publication was paused for a while in late 2020 and early 2021, while the methods were modified). These are also shown in the upper panel of figure \ref{ir-eng}. 

Both direct incidence reconstructions align with the death deconvolution approach, albeit both suggesting slightly earlier peaks before each lockdown. There are two possible explanations for these timing mismatches. It could be that the modelled infection to death distributions increase too rapidly at low durations, with the true fatal disease duration distributions being slightly more right shifted. A less speculative explanation is that deaths occur overwhelmingly in older more vulnerable people, who tend to have lower contact rates and are likely to have reduced these disproportionately relative to the younger healthier population. Sub-populations with lower contact rates peak later than those with higher contact rates, for the reasons discussed in section \ref{models}. So the shift may simply relate to the difference in peak timing expected in a sub-population with lower contact rates.

\begin{figure}
\eps{-90}{.543}{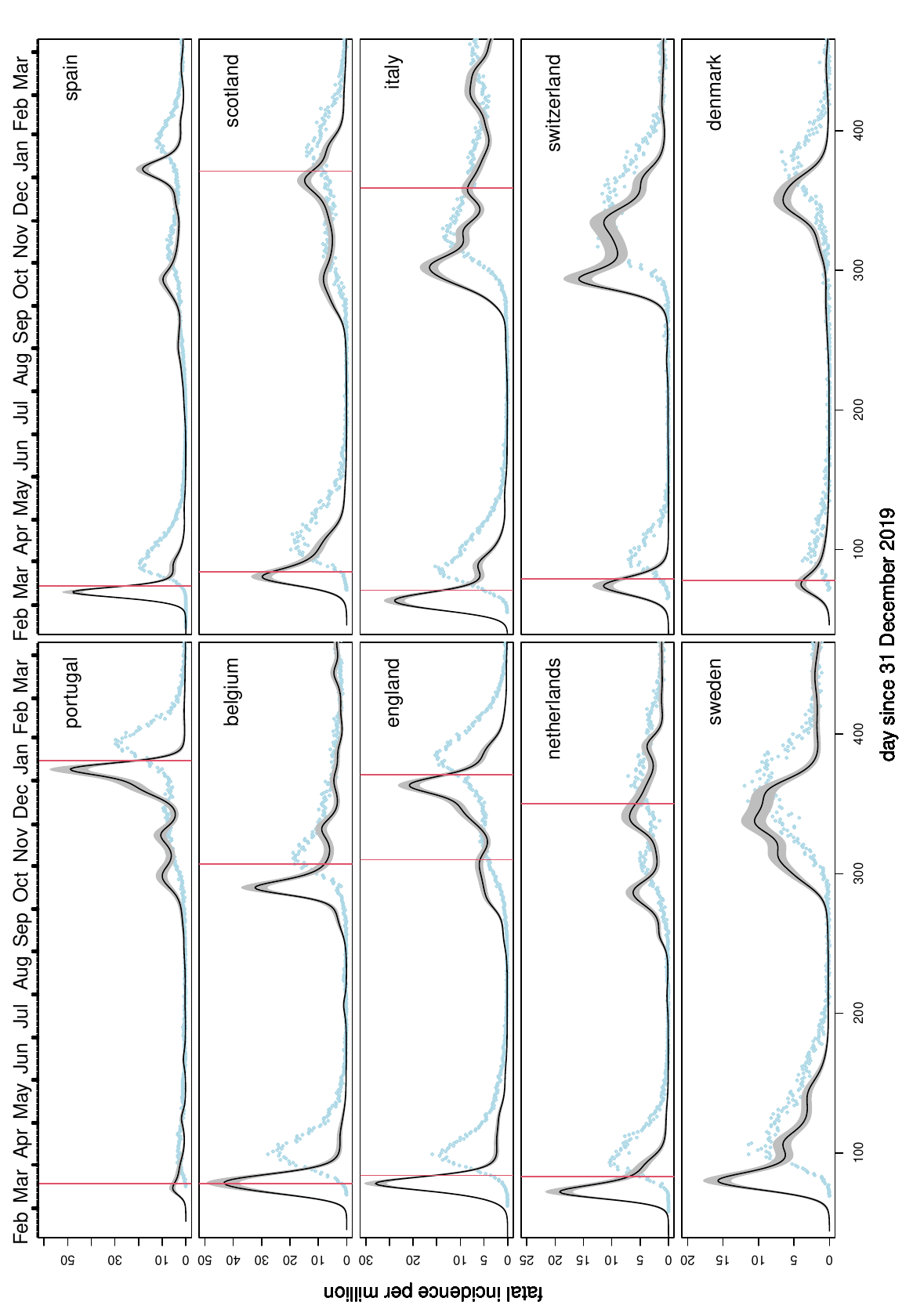}
\caption{Reconstructed fatal incidence for the 10 countries for which reliable daily Covid deaths data by exact day of death were available. Grey confidence bands show reconstructed fatal incidence per million population, with blue circles being the daily deaths per million from which they are obtained. Vertical red lines mark the first day of full national stay at home lockdowns. \label{ifinc}}
\end{figure}

\subsection{International comparisons}

The correspondence between the direct reconstructions of incidence and the deconvolution of daily deaths strongly suggests that the deconvolution approach is sufficiently reliable to be applied to other countries for which daily death data are available by exact day of death (but surveillance surveys are not). In addition to England we were able to obtain data for Belgium, Denmark, Italy, the Netherlands, Portugal, Scotland, Spain, Sweden and Switzerland. The last two are interesting. Both introduced restrictions, but Sweden never introduced full stay at home lockdowns, while Switzerland imposed a first lockdown in March 2020, but thereafter remained substantially more open than its neighbours.

The results are shown in figure \ref{ifinc}. Only for the first Belgian and second Italian lockdowns does the turn-around in infections coincide with lockdown. For waves at other times and/or locations the peak in infections precedes lockdown or decline begins without a full lockdown. Although the results imply that the full lockdowns were largely unnecessary for turning around the waves of infection, the reconstructions are consistent with lockdowns having further suppressed infections, causing infection waves to subside more quickly than might otherwise have occurred. In particular Sweden and Switzerland both experienced broader waves with multiple subsidiary peaks when they did not lock down, a pattern also evident in Italy in the long run up to its eventual January 2021 lockdown. This suppression of infections is also interesting in the light of early model predictions that greater suppression of the first wave would delay rather than prevent infections, leading to larger second waves. Such an effect is certainly consistent with the patterns seen for Portugal, Denmark and Switzerland, and in fact also for Eastern European countries where lockdowns occurred early in the first wave \citep[see e.g.][Chart 4C]{IFA2023-excess}. From this perspective the decision to continue suppression well into the summer of 2020 does not seem optimal in terms of health service loading. See supplementary material for lockdown dates and data sources.

\begin{figure}
\eps{-90}{.543}{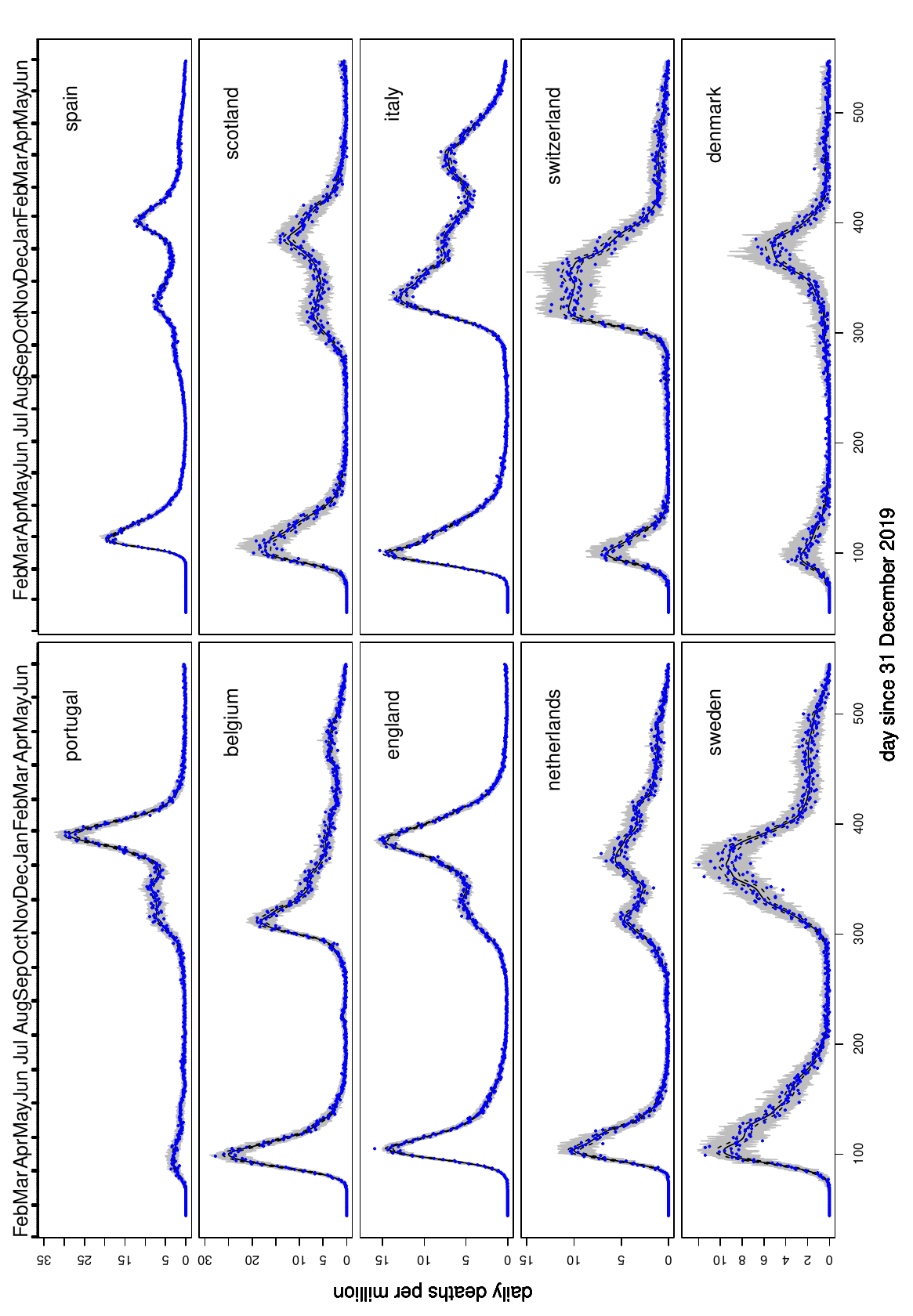}
\caption{Checking plots. Each inferred fatal infection was randomly assigned a duration from the fatal disease duration distribution, to produce a daily death curve implied by the fitted model. This simulation process was repeated 100 times to give the grey curves, which can be compared to the observed daily death data plotted in blue.  \label{sanity}}
\end{figure}

Careful model checking is obviously required when attempting incidence reconstruction. Figure \ref{sanity} shows the results of repeated `forward simulations', in which each inferred fatal infection is randomly assigned a duration from the fatal disease duration distribution. This process results in the simulated daily death rates shown as the collection of grey curves on each plot. Overlaid as blue circles are the original raw daily death data, which should look like a plausible draw from the grey curves if the reconstruction is reasonable. Also overlaid are a simple smooth model fit to the daily deaths with CI for the mean. The plots are unproblematic. Note that models were also tried in which $f$ was represented by an adaptive smooth with time varying smoothness, however these models showed systematic evidence of moderate oversmoothing, presumably related to the rather limited information from which to estimate the several smoothing parameters required. As an illustration of the importance of such model checking, note that at least one group advising UK policymakers attempted to infer incidence  by moving each death back in time according to a random draw from the infection to death distribution. The approach is fundamentally flawed as disease duration is not independent of time of death (e.g. at the start of the  epidemic, deaths are predominantly from people who had short duration diseases). Forward simulation checks of such a method immediately indicate a problem.

\cite{wood2020arxiv,wood2021peakTiming} also included detailed checking of the possibility that the smoothness assumptions in the model might cause mistiming in the presence of surging incidence followed by a lockdown induced collapse -- the dominant narrative when the work was undertaken (although now undermined by REACT-2 results). The checking suggested that the timing results were robust.

\begin{figure}
\eps{-90}{.543}{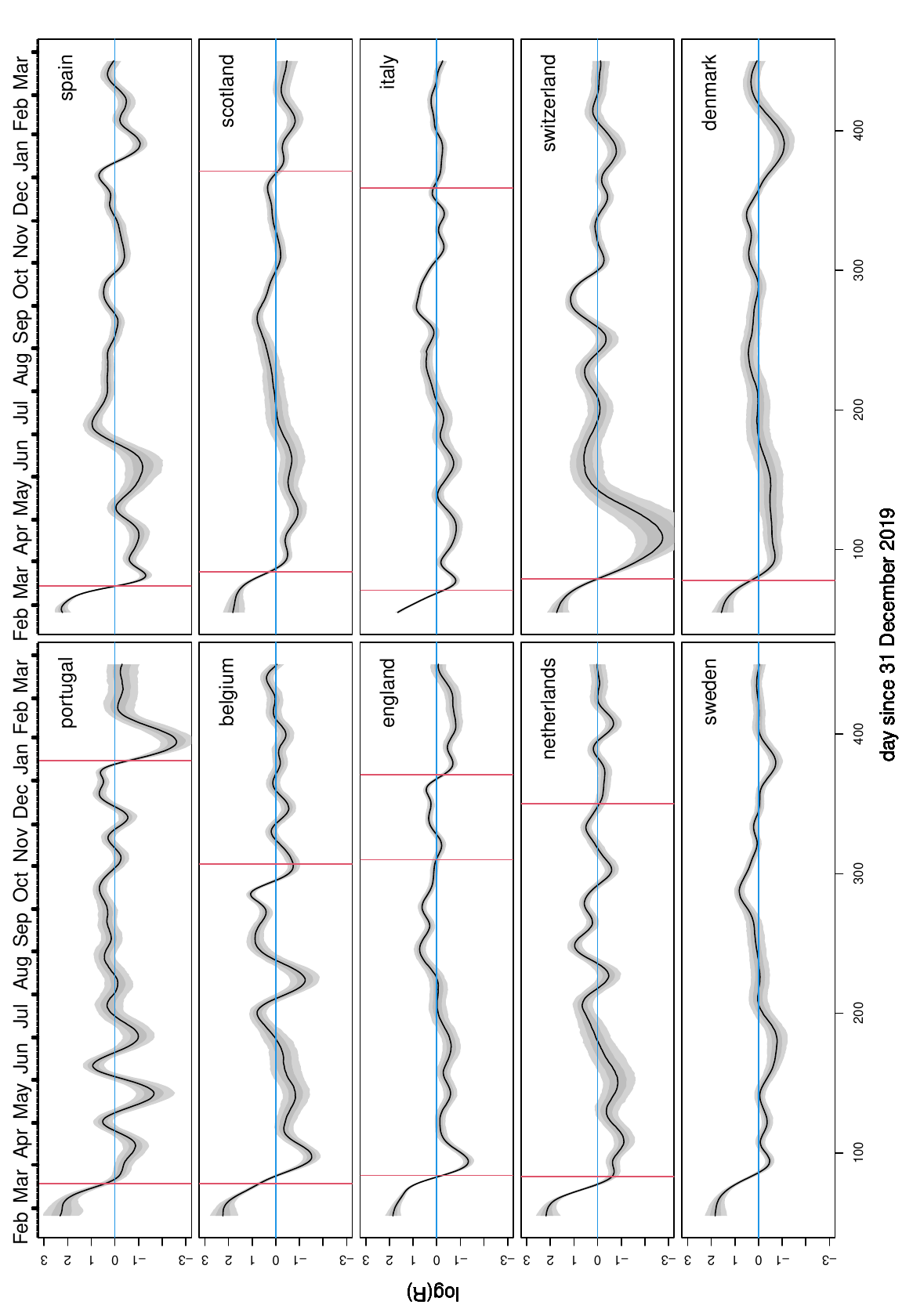}
\caption{The $\log(R)$ trajectory required for a simple SEIR model to produce the reconstructed incidence curves, assuming a mean time to infectivity of 3 days and a mean infectious duration of 5 days. Again vertical red lines mark the lockdown dates. The reconstructions are conditional on an SEIR model structure being reasonable, which it certainly is not at very low incidence rates: hence rapid fluctuations during periods of low incidence are unlikely to be meaningful.    \label{logR}}
\end{figure}

If one is prepared to accept a simple SEIR model as adequate to describe the aggregate epidemic dynamics in a country, then the incidence reconstructions can be converted to equivalent $R$ reconstructions, as shown in figure \ref{logR}. There appear to be no cases for which $R$ had not already declined sharply before lockdown. Only before the first Belgian lockdown was $R$ still appreciably higher than 1, while the second Belgian lockdown apparently came into force when $R$ was already at a low point not seen subsequently. In contrast, at the first lockdown the Netherlands already had $R$ well below 1, but otherwise $R$ was typically around 1 at each country's lockdowns. Only England, Italy and the Netherlands have $R<1$ clearly before the first lockdown.    

\subsection{The alternative lockdown narrative \label{alt-ld}}

\begin{figure}
\eps{0}{.545}{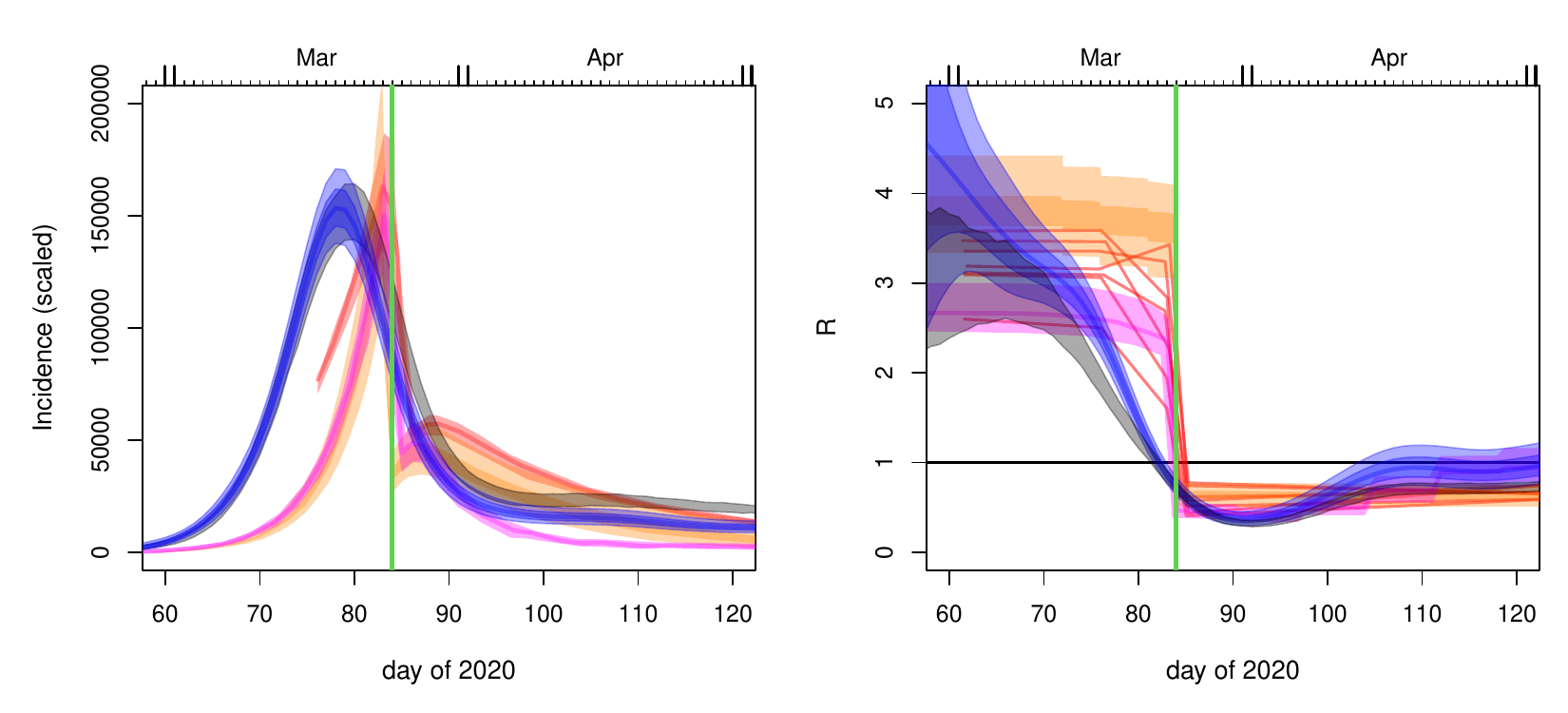}

\vspace*{-.5cm}

\caption{Left: incidence reconstructions (vertically scaled for plotting). Right: $R$  reconstructions. The green vertical line is the first day of lockdown. Red: IC \cite{rep41} ($R$ separate for each region); Orange: IC \cite{flaxman2020lockdown}; Pink: MRC for London \cite{Birrell2021}; Grey: replication of \cite{rep41} by \cite{woodwit2021plos} relaxing restrictive assumptions on $R$; Blue: replication of \cite{flaxman2020lockdown} from \cite{wood2021peakTiming}, relaxing the restrictive assumptions on $R$. \label{alt-inc}}
\end{figure}

In the UK, analyses from Imperial College \citep{flaxman2020lockdown,rep41,knock21}, and the MRC unit in Cambridge 
\citep{Birrell2021} were widely covered and highly influential in promoting the idea that lockdown was the essential component in turning around the first wave of infection. The analyses fitted epidemic models to daily death data, and to other clinical data streams in the case of \cite{rep41,knock21}. All apparently showed surging incidence up until the eve of the first lockdown, as shown by the red, orange and pink bands in figure \ref{alt-inc}.

The \cite{flaxman2020lockdown} paper attempted to fit a simple renewal model to death data from multiple European countries, assuming that different NPIs had the same multiplicative affect on transmission in each country, irrespective of their order of application, except for the full lockdown effect, which was allowed more country to country variability. To allow for the fact that Sweden did not lock down, the final intervention in Sweden was modelled as if it was lockdown. The approach has been widely criticised \citep[see e.g.][]{chin2021flaxman}. A particularly insidious problem is the model's treatment of $R$ after full lockdown: $R $ was modelled as a step function, changing only when government policy changed, so constant after lockdown. However, the basic statistical reasoning detailed in section \ref{models2} shows that the average $R$ can not be constant after lockdown, if lockdown is effective at reducing transmission rates. Instead, after an initial decline a recovery is expected. Exactly this effect is seen in figure \ref{ir-eng}, but is precluded by the analysis model of \cite{flaxman2020lockdown}. It is difficult to reason intuitively about the consequences of such a structural problem for a highly non-linear model, so \citet[][see also Appendix \ref{smoothing}]{wood2021peakTiming} re-implemented the \cite{flaxman2020lockdown} model for England, with the restrictive step function replaced by a cubic spline model for $\log(R)$. The results then match figure \ref{ir-eng}, as the blue bands in figure \ref{alt-inc} show.

\cite{Birrell2021} also reported surging incidence up until lockdown (pink in figure \ref{alt-inc}), based on an epidemic model fitted to death data. In this case $R$ was controlled by a contact rate modifier step function with weekly steps, {\em except in the period before lockdown, where it was constant}. In other words, increasing incidence until lockdown was simply built into the model.

\cite{rep41} fitted an age structured multi-compartment model to health service death and hospital occupancy data, alongside PCR and antibody testing data. The model had some 700 state variables, but inference employed particle filtering with only 96 particles \citep[doubled for the eventually published][]{knock21}. For the 7 English health service regions they again purported to show that incidence was increasing and $R>1$ right up to the eve of the first lockdown (red in figure \ref{alt-inc}). In this case $R$ was controlled by a piecewise linear contact rate modifier with 12 knots at selected government intervention points. Again there is insufficient flexibility to capture the post lockdown dip and recovery in $R$ expected if lockdown reduces contact rates. \citet[][see also Appendix \ref{smoothing}]{woodwit2021plos} replicated the analysis, replacing the contact rate modifier with an adaptive spline and resetting several rate constants to the values given in the literature cited by \cite{rep41} as their source. A simpler model estimation scheme was used in place of particle filtering. Again, on relaxation of the strong and unrealistic assumptions on contact rate changes, the results aligned with figure \ref{ir-eng}, as the grey bands in figure \ref{alt-inc} show.

The other major plank of the narrative of lockdown necessity, was the fact that mathematical models had predicted that lockdown was essential to turn around infection waves, and that after lockdowns were imposed infections indeed declined. The models were not validated for prediction in advance, and the fact that they were able to predict that Covid spread could be massively reduced by suppressing  human contact to the maximum extent possible is an especially undemanding check of model sanity. 

Two more discerning tests were available. The first one consists of forward mortality predictions.  Early in 2020 Imperial College published a study giving the number of predicted deaths likely to accrue under different social distancing scenarios for a number of countries. \cite{report12} predicted about 35,000 first wave Covid deaths for Sweden under the ‘social distancing of the whole population’ scenario, short of full lockdown, which is the closest scenario to what Sweden actually did. This is interesting as it represented the only first wave test of the models’ ability to predict what might happen without lockdown. Sweden in fact experienced fewer than 6,000 first wave deaths.

\begin{figure}
\eps{0}{.5}{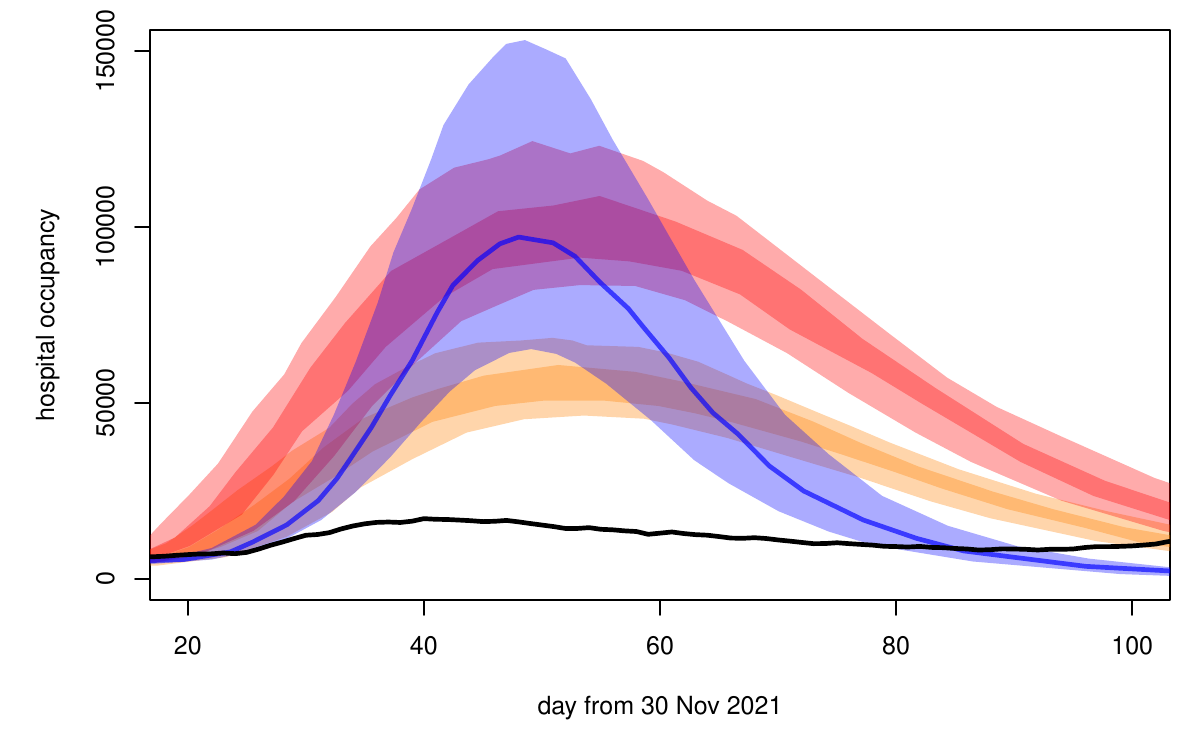}

\vspace*{-.5cm}

\caption{LSHTM projections for NHS England Covid hospital bed occupancy as a result of omicron, without lockdown in orange and red. These are the scenarios that are most optimistic about booster efficacy, with the orange representing low vaccine escape assumptions and the red high vaccine escape. In blue are Warwick projections assuming omicron to be 50\% as severe as delta. This is the modelling used in the 19th December SPI-M advice. The black line is actual occupancy. Scenarios digitized from the source documents. Actual occupancy figures are NHS England data. \label{spi-omi}}
\end{figure}

The second test of the models' ability to predict what would happen in the absence of lockdowns came with the omicron variant at the end of 2021. The UK government's Scientific Advisory Group for Emergencies (SAGE) relied heavily on the SPI-M committee which synthesised modelling work on Covid to advise on policy. On 19th December 2021 it issued advice on the omicron variant \citep{SPI-M-omicron}, strongly suggesting the urgent need for a fourth lockdown. The following edited extract from the summary gives a flavour of the advice.
\begin{quote}
\ldots A key consideration for decision
making is how to avert unsustainable pressure on health and care settings\ldots
If the coming wave rises comparatively slowly, then a short intervention for, say, a few
weeks can prolong the wave’s duration and reduce its peak so that admissions and
hospital occupancy remain below levels that would compromise quality of care \ldots enacting an intervention early would give time to detect whether such an
intervention is insufficient to avoid a compromise of quality of care and adjust accordingly.
If measures are implemented only later \ldots measures would need to be in place for longer and might be too late to
avert very high admissions\ldots
\end{quote}
The detailed advice was based on modelling from the London School of Hygiene and Tropical Medicine \citep{barnard2021omicron} and the University of Warwick \citep{keeling2021omicron}, with the former given substantially more prominence. The government declined to lock down again, so it is possible to compare model projections with reality. Figure \ref{spi-omi} does this. In fact the Warwick modelling, as well as presenting a scenario much worse than the one shown, also showed a scenario under the assumption that omicron was only 10\% as severe as the delta variant. The lower part of the interval for this scenario does include what actually happened, but it is fair to say that the SPI-M advice did not present this scenario as one that was credible.
\section{Discussion}

The response to Covid was extraordinary in the extent to which it took place online. Initially in the intense pressure in favour of locking down that built on social media, and then in the movement of so much human and scientific interaction online, once lockdowns and other social distancing measures had been implemented. The tendency for online interactions to polarize, amplify exaggeration, and rapidly promote fashions of thought, panics and enthusiasms, while encouraging availability and confirmation bias \citep[e.g.][]{kahneman2011tfs},  has been well-documented by social commentators \citep[e.g.][]{ascap}. These tendencies probably make the online world a less than ideal forum for the careful weighing of evidence. They may also serve to promote an adversarial approach to scientific questions, in which the scientist acts as an advocate, whose role is to marshal the data and arguments supporting their theory, rather than as a more neutral interrogator of what data may reveal about reality. The adversarial approach may have advantages, when time is not pressing and there are opposing advocates to attempt falsification, but is perhaps less suited to an emergency, especially if opposing views are characterized as presenting a danger to public health, for example. 

The question of lockdown's necessity in turning around waves of infection provides an example where the most careful evidence weighing was appropriate, given the profoundly damaging nature of the intervention. Whatever one's views about how risks should have been balanced in the initial decision to lock down, there was surely an urgent need for rapid and clear eyed evaluation of whether this experimental intervention had in fact been necessary, as soon as possible after its imposition. Instead models appear to have been treated as evidence
and informal intuitive reasoning preferred to what the data strongly implied. In part the reliance on models may reflect a confusion between updating beliefs and validating them: the notion that updating the distribution of a model's parameters using data in some way validates the model, which it does only in the limited sense of failing to immediately falsify it. The models appearing to indicate the necessity of lockdown can indeed be updated/fitted using data, but highly non-linear models often have the beguiling flexibility to be able to reproduce a wide range of data, quite irrespective of how their structure reflects the real data generating mechanism. One could argue that the media's response to this issue also tended to confuse the majority opinion of scientists with scientific evidence, emphasising perhaps that expertise is a good reason for listening carefully to an expert's argument, but not for accepting it.

The detachment from external objective reality promoted by the online environment provides fertile ground for the post-modern idea that language in fact creates, controls or {\em is} reality, at least in the social sphere, perhaps feeding the cognate notion that mathematics, the language of science, is equivalent to or controls scientific reality. The `illusion of control' that this creates \citep[e.g.][]{gupta2001} is seductive, but unhelpful if it results in an excessive effort being devoted to modelling rather than measurement. This is not an argument against quantitative science. For example, the scientifically advised central UK government did not fall for the deeply unscientific and innumerate belief \citep[see][]{dahlem1998} that an endemic disease could, for the first time in history, be eliminated by physical distancing measures, if only these were sufficiently stringent and prolonged (`zero Covid'). Sticking to the quantitative science, in this respect, avoided the even greater collateral damage of harsher more prolonged measures.

But there was room for improvement in the balance between theory and data, and between modelling and measurement. The public availability of data was in many respects exemplary, with the ONS  providing solid evidence on many aspects of the crisis, along with studies such as ISARIC and REACT. But other data was effectively closed to general scrutiny. Data on nosocomial infection was particularly closely guarded, and sensitivity over this appeared to also limit the availability of some other data, such as that relating to the time from first symptoms to death. The ability to independently check and replicate the modelling used to advice policy is severely limited if such data are restricted to an inner circle of advisors. Academic statisticians can not do their job, of thinking critically about data, unless the data are accessible. Similarly, while some modellers, such as the Imperial College group, took repeatability seriously enough that replication was possible (providing data, comprehensive statement of models and code), other models used by SPI-M were impractical to replicate given what was provided. 

Another obvious area of concern is the length of time that it took for randomized surveillance sampling to get underway, given that PCR tests were available from January 2020, when the first UK Covid cases were confirmed (UK government figures put PCR test processing capacity at over 6000 per day by March 20th, up from 1500 per day on March 11th. With standard statistical methods for batch testing, a fraction of that capacity would suffice for a useful survey, and the labour force survey sampling frame was already available). In a situation so serious that almost the whole population could be confined to their own homes for 23 hours per day without external in person contact, it seems incongruous that the first surveillance samples to {\em measure} the actual state of the epidemic were not taken until 25th April 2020, nearly 3 months after the first UK Covid cases, 2 months after Lombardy and 7 weeks after the first UK death. It seems unlikely that deficiencies in a rapidly rolled out survey, refined as it progressed, could have been worse than not having surveillance data.

A further problem is the limited role that the collateral risks from the measures seem to have played in decision making. At the very least a `red-team' with similar heft to SAGE would seem appropriate as soon as a massively costly experimental intervention becomes a serious possibility. Such a team, might, for example, have questioned the fairness of allowing the cost per life year saved from Covid to be many times the usual NICE threshold  for approval of an intervention (about \pounds 30,000 per QALY). 

Some aspects of the combination of evidence also appear to have been less than ideal. For example the 19th December 2021 SPI-M statement on omicron (see section \ref{alt-ld}) creates the worrying impression that the Warwick modelling may have been somewhat down-weighted in the advice because of the very wide range of outcomes that it suggested were possible, with the LSHTM model given more prominence as a result of its apparently lower uncertainty. But the LSHTM model's increased precision was an illusion created in large part by neglecting the very wide uncertainty in the relationship between vaccine efficacy at blocking transmission and hospitalization used in the modelling. 

Also worth discussion is the SPI-M insistence that they were presenting projections (or scenarios), but not predictions, while at the same time making probability statements about them. This seems at best philosophically awkward. How is a policy maker to interpret a probability statement about a projection if it is not a prediction? Perhaps `if the world is sufficiently  like the model then this is the probability of the event of interest'? To act on such a probability the policy maker then needs to have some reasonable idea of the probability that the world is like the model. But to declare that projections are not predictions is to declare that this latter probability is unknown. We think that the resolution of this problem probably leads back to the need to statistically validate models for prediction. Otherwise basing health interventions on model predictions seems worryingly close to licensing a new drug without a clinical trial.

Finally, what was the strategy for dealing with the eventuality that an effective vaccine could not be developed? Without such a strategy it is difficult to see that risk was in fact being managed, however matters eventually turned out. If there was a strategy, but it was not publicly discussed, it is difficult to see it as having the democratic legitimacy one might expect in an open society.


\vspace{-.2cm}

\subsection*{Acknowledgments}

We are especially grateful to Nancy Reid for suggesting that this paper be written. It would not have been otherwise. Thanks also to Jonathan Rougier, Peter Green, Nicole Augustin, Matteo Fasiolo, Helen Colhoun and Dan Coggan for various discussions of some of the issues raised here, and to the referees for a wealth of useful comments and for providing a number of extra references. SNW is also grateful for help from and discussions with a number of medics who cannot be named, and Dr Alistair Montgomery who can be. 


\begin{appendix}
\section{Iterating life table demography \label{demog}}

This appendix provides some details on the iteration of ageing and deaths used in section \ref{sec:excess}.

Let $y_i$ denote the deaths in week $w_i$ of the year, corresponding to time $t_i$ since the start of the data. To estimate the annual cycle in death rates the generalized additive model
$$
\mu_i = f_1(w_i) + f_2(t_i),~~~~~~ \frac{y_i - \mu_i}{\sigma} \sim t_\nu
$$
was estimated from 2017-19 data, where $f_1$ is a cyclic smooth function and $f_2$ is a centred slowly varying smooth function, while $\sigma$ and $\nu$ are parameters to be estimated. Then $d_w = \hat f_1(w)/\sum_{w=1}^{52} \hat f_1(w)$ defines the multiplier of average weekly mortality required to account for seasonal variation. 

The UK population at the start of 2017 or 2020 is available in 1 year age classes, from 0 to 99 plus a `100+' class. All one year age class populations were then split into 52 one week age classes. This was done by fitting a monotonic interpolating spline to the annual cumulative population by age data, and then simply differencing the resulting fit to obtain weekly populations. In this way the weekly populations vary smoothly, without year end discontinuity, while the total for each year exactly matches the original yearly data. The approach neglects seasonal birth rate fluctuations. The population in the first week age class is also taken to be the weekly birth rate (the crudeness of this approximation having negligible impact on total deaths). Note that the method also works with data aggregated more coarsely than by yearly age classes.

The life tables provide instantaneous per capita death rates $m_a$ (units year$^{-1}$) for each one year age group $a=0,\ldots, 100$. The average proportion of the age group then dying in one week is $q_a = 1-\exp(-m_a/52)$. Hence the proportion of one year age group $a$ dying in week of year $w$ is  $q_a d_w$. Given these preliminaries, the demography is iterated forward using a weekly time step in which the expected deaths are subtracted from the population in each weekly age class before each class is shifted onwards one week, and new births are added to the first age class. The per capita mortality rate in a weekly age class is taken as the mortality in its corresponding yearly age class. 

When iterated for 3 years from the estimated population by age at the start of 2017 this approach slightly underestimates actual deaths  by just under 50 out of 1.8 million ($<0.0026\%$). This slight underestimation will lead to a slight overestimation in excess deaths (somewhere around $0.05\%$). Code and data used are provided in the supplementary material.

\section{Modelling with smooth functions \label{smoothing}}
Both the death deconvolution models and the replications of \cite{flaxman2020lockdown} and \cite{rep41}, covered in section \ref{lockdown}, are statistical models in which smooth functions of time are to be estimated alongside other parameters. A basis expansion is employed for the smooth function, $f(t) = \sum_{k=1}^K \beta_k b_k(t)$ where $\beta_k$ is an unknown coefficient targeted by statistical inference and $b_k(t)$ a known basis function chosen for good approximation theoretic properties. A cubic spline basis is convenient. Associated with $f(t)$ is a smoothing penalty, such as $\lambda \int f^{\prime\prime}(t)^2 dt = \lambda \bp\ts {\bf S}\bp$ ($\bf S$ known), which can be used to penalize complexity of $f$ during inference, tuneably via the smoothing parameter $\lambda$. In a Bayesian setting it is natural to view such a penalty as being induced by an improper Gaussian smoothing prior $\bp \sim N({\bf 0}, {\bf S}^-/\lambda)$. 

Denoting the model log likelihood as $l$ and expanding $\bp$ to include any other model parameters (and zero padding $\bf S$ accordingly), then the maximum penalized likelihood estimates of $\bp$ are given by
\beq
\hat \bp = \underset{\bp}{\text{argmax}} ~ l(\bp) - \frac{\lambda}{2} \bp \ts {\bf S} \bp, \label{preg}
\eeq  
which is also the posterior mode under the Bayesian view. Pushing the Bayesian view further gives the large sample approximation
$$
\bp | {\bf y} \sim N(\hat \bp, {\bf V}_\beta),
$$ 
where ${\bf V}_\beta = (-\ilpddif{l}{\bp}{\bp \ts} + \lambda {\bf S})^{-1}$. Writing $\pi_G(\bp|{\bf y})$ for this Gaussian approximation to the posterior and  $\pi(\bp)$ for the smoothing prior, the marginal likelihood is approximately $\exp\{l(\hat \bp)\}\pi(\hat \bp)/\pi_G(\hat \bp|{\bf y})$ (Laplace approximation) which can be maximized to estimate $\lambda$. This approach is equally applicable when there are several smoothing parameters and a penalty of the form $\bp \ts {\bf S}_\lambda \bp$ where ${\bf S}_\lambda = \sum \lambda_j {\bf S}_j$.

Direct approximate marginal likelihood maximisation would involve nested optimization, which can be tedious to implement for a bespoke dynamic model. However \cite{wood2016gfs} demonstrate how it can be approximately optimized using a simple iteration that alternates Newton optimization of (\ref{preg}) with updates
$$
\lambda_j \leftarrow \frac{\text{tr}({\bf S}_\lambda^-{\bf S}_j) - \text{tr}({\bf V}_\beta {\bf S}_j)}{\hat \bp\ts {\bf S}_j\hat \bp} \lambda_j.
$$
(note that for many penalties $\text{tr}({\bf S}_\lambda^-{\bf S}_j) = \text{rank}({\bf S}_j)/\lambda_j$.) Hence inference about $\bm \lambda$ and $\bp$ then requires only first and second derivatives of the model log likelihood with respect to $\bp$. For the deconvolution model these are straightforward to obtain. For the renewal model of \cite{flaxman2020lockdown} an iterative system for the first and second derivatives of the (discrete time) dynamic model is required, but is relatively straightforward to produce. 

For the \cite{rep41} ODE model a system of `sensitivity' ODEs has to be solved in order to compute first derivatives - this involves 10s of thousands of ODEs, which, while inconvenient, turns out to be numerically inexpensive. However the second derivative system is impractical. Fortunately there is an alternative. Solving (\ref{preg}) by quasi-Newton, the Hessian of the log likelihood required to compute ${\bf V}_\beta$ can be obtained by finite differencing the numerically exact first derivatives, in which case the smoothing parameter updates can proceed using the above update formula, with the approximate posterior available `for free'.

\end{appendix}


\begin{center}
{\LARGE \bf Authors' reply to the Discussion of `Some statistical aspects of the Covid-19 response'}\\

\bigskip

Simon N Wood \\ School of Mathematics, University of Edinburgh, Edinburgh UK.\\
Ernst C. Wit\\ Institute of Computing, Univerit\`a della Svizzera italiana, {Lugano}, {Switzerland}
\\
Paul M. McKeigue\\ 
College of Medicine and Veterinary Medicine, {University of
Edinburgh}, Edinburgh {UK}\footnote{The four student authors could not be involved in preparing the response, because of the short time available, and the fact that in the lengthy period since the paper was written they have all moved on to other employment. }
\end{center}

\setcounter{section}{0}
\setcounter{subsection}{0}
\renewcommand\thesection{\arabic{section}}

\noindent Discussion contributors were given 2 months to finalize their contributions, after which they were passed to us to respond to in about 4 weeks. We were told that this timescale was to ensure publication in 2025. We met the deadline, the proofs were returned in July 2025, but the paper was not published in 2025. As of late November 2025, the contributions of Birrell and de Angelis are not available on the JRSSA website: we have provided the links to the other discussion contributions in our response below. 

\section{General response}

\begin{quote}
{\em \ldots humans find it very difficult to consciously reflect on a large number of datapoints and weigh them against each other \ldots That's why when faced by complex issues -- whether a loan request, a pandemic or a war -- we often seek a single reason to take a particular course of action and ignore all other considerations. This is the fallacy of the single cause.} \\  Yuval Noah Hari {\em Nexus}.
\end{quote}

What went well in the UK Covid response? This question, posed by Jose and others, is a good place to start, given our paper's explicit focus on what did {\em not} go so well. Here are {\em some} of what we believe are scientific highlights. The ONS survey, {\em measuring} the state of the epidemic was internationally leading: why it was not widely replicated internationally is a mystery. Similarly REACT and REACT2 and the openness of much other data were exemplary. The central UK government not pursuing the pseudoscience of zero Covid and, perhaps through growing model scepticism, declining to lock down for Omicron were also huge positives. The UK's lab based biomedical scientists, like those of several other countries, obviously did a great job on vaccines.

We would like to thank all the people who have taken the time to comment on a paper that for many would have been much more comfortable to ignore --- although it is a pity that Sir Chris Whitty, Dame Angela McLean, Sir Ian Diamond, Neil Ferguson OBE and Grahame Medley OBE were too busy (or otherwise unwilling) to contribute when invited to do so. Those closest to the decision processes will surely have had the most coherent counter arguments to what we have written and it would have been good for those to be public. Also absent from the discussion are the trenchant criticisms of the UK Covid response that a number of respected statisticians are willing to make in private. We think that it is time to make those public, although we can well appreciate the reluctance to do so.  

In responding, we begin by noting that none of the critical comments directly addresses the central issue of our paper: that every Covid intervention involved not only benefits, but also significant costs. Based on pre-existing data and prior research, it should have been apparent, early in the pandemic, that the long term health (and other) risks stemming from the economic consequences of the measures were comparable to -- or even greater than -- the health risks those measures aimed to mitigate. 

Effectively ignoring these trade-offs and endorsing a maximal remediation strategy required an appeal  to an absolute imperative: in the UK, avoiding the collapse of the NHS\footnote{what this meant was not spelled out -- perhaps being forced into a mode of operation normally seen only in war zones?}. The unacceptability of that scenario apparently justified virtually any future cost. However, such an absolutist rationale is only defensible if the maximal measures were indeed {\em necessary} to prevent the outcome, and the most reliable reconstructions of incidence suggest that this was probably not the case. Without this, the justification for risk-distorting fear-messaging, the suppression of dissent as misinformation, the up-ending of usual evidence norms in science, the suspension of civil liberties and the damage done to the young in the interests of the old, are even more difficult to view as sound, evidence based, public health measures.

Where did this approach come from, of near zero risk appetite in the short term combined with enormous risk appetite over longer time frames? The latter, at least, is somewhat consistent with the response to climate change. It is scientifically clear that exacerbating climate change will come with truly existential medium to long term risks. Nevertheless, whereas the UK Covid public spending alone of \pounds 310--410 Billion has received rather limited scrutiny, a planned UK spending package of \pounds 27 Billion, to tackle climate change, was canceled in 2024 as now \emph{unaffordable}.  In this context, it is also concerning that poor scientific messaging and modelling around Covid has fueled somewhat justifiable public mistrust, including a far less justifiable scepticism about climate messaging and modelling.

Moving on to the many detailed comments raised, the most widespread, legitimate and substantive technical criticism is that we failed to allow for improvements in life expectancy in the iterated life table method for calculating excess deaths, when the steady rise seen pre-Covid made this a reasonable assumption (McDonald, Fisch, Dodd et al., Andrews  and a number of in person comments after the meeting \url{https://doi.org/10.1093/jrsssa/qnaf091}, \url{https://doi.org/10.1093/jrsssa/qnaf112}). This point is completely fair, and very easy to correct using the life table method and supplied code, by simply reducing $d_w$ each year by the factor required to increase life expectancies by the expected amount. Using the pre-Covid figure of 18 days life expectancy increase per year, results in around 28000 more cumulative excess deaths than we calculated in the paper, as can very easily be checked using our supplementary code. The objection was also raised that we accumulated excess deaths from the start of 2020. Given Sheila Bird's point, that the first known UK Covid death eventually turned out to have been on 31st January 2020, we are not sure that this is unreasonable\footnote{the criticism of quoting the later date when discussing ONS first sampling dates seems less reasonable: no one in April 2020 could be blamed for responding too slowly to a death not discovered until  September 2020.}. Nevertheless, if we instead choose to start accumulating from the excess death minimum in 2020, then nearly 6000 excess deaths are added to the total, taking us to around 130 thousand. The discrepancy with methods that do not fully account for an ageing population (or life expectancy improvement) is still very large: our figure 5 shows why this has to be the case.      

\begin{figure}
\vspace*{-1cm}
\eps{0}{.55}{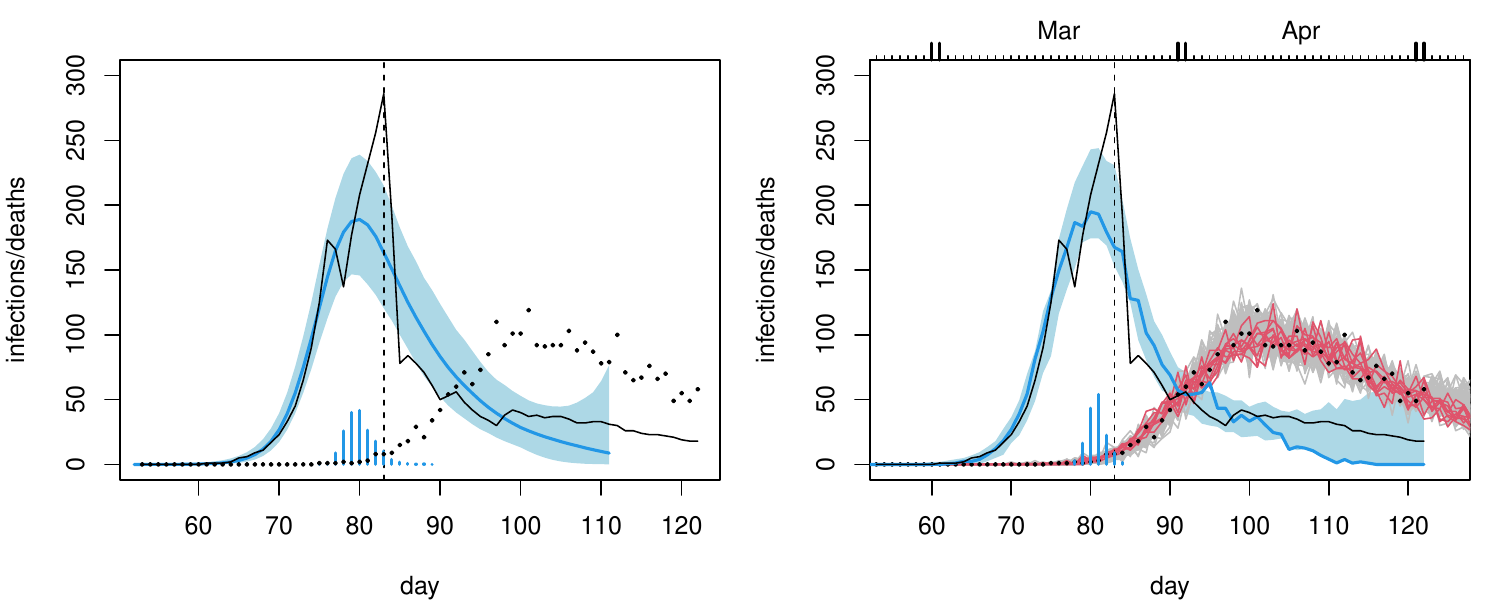}
\vspace*{-.4cm}

\caption{Left: Replication of the Birrell et al. simulation shown in their lower panel figure 2, using their assumed incidence profile, but also showing the uncertainty in peak location for the \cite{wood2021peaktiming} reconstruction shown. The blue bar chart shows the posterior distribution of the peak location (scaled), which clearly includes the truth. Right: equivalent reconstruction using the simulation based deconvolution method of \cite{wood25pt2}, which does not make a smoothness assumption (red trajectories are replicate death trajectories simulated from the dark blue incidence reconstruction, grey are bootstrap replicates of the same). Again there is insufficient information in the signal to resolve the high frequency feature of the simulated incidence, but in this case the misstiming is reduced, while the distribution of inferred peak time again includes the truth. \label{birrell-fig} }
\end{figure}

\begin{figure}
\vspace*{-1.1cm}
\eps{0}{.6}{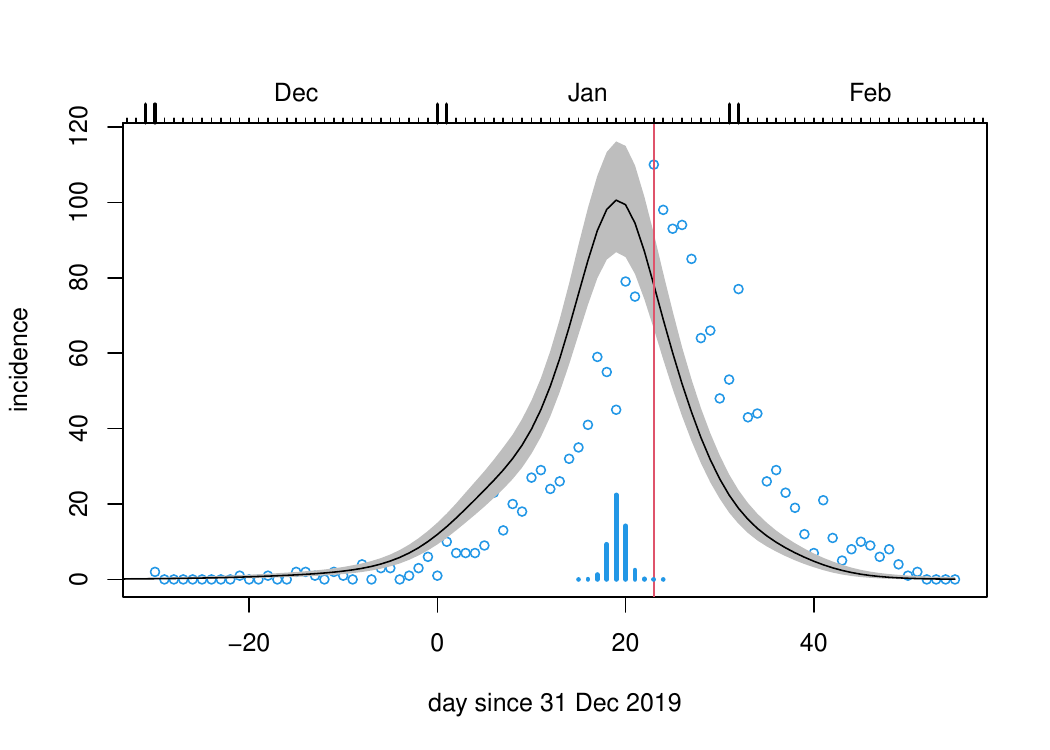}
\vspace*{-.5cm}

\caption{Wuhan fatal incidence reconstructed by deconvolution (black and grey) from the recorded time of first symptom onset for fatal Covid cases. Blue circles are the number of fatal cases with onset on the given day \citep[China CDC data,][]{bai2020wuhan}. The red line is at the date of Wuhan lockdown on 23 January. The blue bar chart is the simulated posterior distribution of time of peak incidence (scaled) corresponding to a 95\% CI of 17th to 21st of January 2020.
\label{wuhan-fig} }
\end{figure}

A second important objection to the work comes from Paul Birrell, Lorenzo Pellis, Robert Verity and their colleagues. That is the concern that the infection peak timing results could be smoothing artefacts if the real incidence was in fact exponential increase until lockdown followed by collapse. \cite{wood2021peaktiming} examined this question in detail (as did the paper's arXiv version from June 2020), covering exactly the point Birrell, Verity et al. discuss and demonstrating that the effect is insufficient to explain the results on the real data. \cite{wood25pt2} also provides a method that eliminates the smoothness assumption and provides a statistical test for hypothetical incidence profiles: the timing results do not change and, for England and Wales, a profile of exponential incidence growth with collapse at lockdown is firmly rejected. In any case, such an inference artefact cannot apply to the direct sampling based estimates from REACT2 and the ONS (or the direct estimates from Wuhan, in figure \ref{wuhan-fig}, showing a similar pattern). 

Several discussants also provide simulations intended to support their point. The Pellis simulations do not really seem relevant. His figure 1 shows the fit of an obviously incorrect smooth model, from which incorrect $R$ estimates are then derived: what else is to be expected? Furthermore, any such misfit can easily be identified from appropriate residual plots. The peak displacements seen in the Verity et al. simulation (\url{https://doi.org/10.1093/jrsssa/qnaf108}) are consistent with the discussion in  \cite{wood2021peaktiming}, given the slower epidemic dynamics that they consider: their simulated doubling time of 10 days is much longer than that which pertained in the first Covid wave. So, the discussion of this issue in the context of the real data remains unaffected. For one simulation, Birrell gets an incidence reconstruction mistimed by 4 days, based on a truth that increases rapidly, then decreases sharply for 2 days and then increases again for the week until lockdown. Any method will have difficulty resolving such a feature from data obtained by convolving it with the time to death distribution\footnote{for example, how would the MRC method fare, given its assumption of weekly step changes with smoothness imposed by a random walk prior?}, but what Birrell has not shown is the uncertainty of the estimated peak location, which is then high, partly also because of the low numbers in the simulations. Figure \ref{birrell-fig} corrects this omission, while its right panel shows the result of applying \cite{wood25pt2} to the same data. Birrell's regional point echoes \cite{woodwit2021plos}, but the uncertainty his simulations reveal at lower levels of aggregation emphasize the difficulty of separating what is data driven and what model driven from more local data. Actually, it is exactly the intrinsic uncertainty associated with relatively low counts that makes a daily deaths deconvolution for Wuhan too uncertain to be useful, but it is worth noting that Wuhan actually provides another data source for reliable direct incidence reconstruction: first symptom onset date was recorded for all fatal cases\footnote{We are grateful to Piet Streicher for pointing this out to us after the meeting.}. Figure \ref{wuhan-fig} deconvolves these data with the infection to onset distribution to reconstruct incidence (the much narrower distribution greatly eases the deconvolution task).

Fundamentally, the problem with smoothing is not the smoothness {\em assumption} \citep[which can anyway be removed as in ][]{wood25pt2}, but the smoothing {\em operation} inherent in the convolution of incidence with the disease duration distribution. The difficulty of reversing that is what makes it difficult to distinguish the fit of a discontinuous incidence curve from a smooth one (but again see Wood 2025 for a statistical test). Once that is recognized, there is no justification for asserting  that model fitting strongly supports the narrative of surging incidence up until lockdown. It cannot, and as soon as it was demonstrated (in early May 2020) that a continuous incidence curve peaking well before lockdown explains the data  at least as well, there was no excuse for continuing with the assertion. Further, the earlier peak model was consistent with the data that was then available on the extent of contact rate reduction, the drop seen in NHS 111 calls for respiratory diseases and the fact that Sweden's wave had evidently turned within a day or two of the UK, without lockdown. The careful checking of the possibility of a smoothing artefact in \cite{wood2021peaktiming} also suggested that the peak before lockdown result was likely to be real.   

In our view, from the moment in May 2020 that the simple deconvolution 
results first appeared it was not reasonable or responsible to continue arguing that the evidence strongly supported lockdown having caused incidence to decline. It was at least as plausible (more likely given other data) that incidence was in sharp decline before lockdown. Given the evidence that has accumulated since then, is it really reasonable to insist that the question is still unresolved? 

The preceding sentence benefits from hindsight, of course, something that several discussants mention about the paper more generally. We actually think that much of what we discuss suggests poor risk management {\em given what was known at the time}, but of course some things, like the depth of the economic damage, or that `build back better' was an illusion, or the damage to school aged children, are now much clearer than they were in 2020. However, we think that the most significant piece of true hindsight, framing current discussions, is the fact that it proved possible to produce the first vaccines against a corona virus and to do so in a very short space of time\footnote{albeit the vaccines could not produce the herd immunity that was assumed in the discussions of 2020}. At the pandemic outset, it is difficult to see a rational basis for assuming that such a development was more likely than not, let alone certain. Given that significant uncertainty, Sweden's approach of  managing Covid as a part of everyday life appears rational, but the UK's approach less so.

We now turn to the three invited discussants, Goldblatt, DeAngelis and Bird\footnote{the meeting did not follow the Society's usual proposer-seconder tradition.}:  Peter Goldblatt's most substantive point seeks to disconnect the life loss identified in the Marmot report from the economic disruption caused by the 2008 crisis.  Fundamental to this point is a belief that austerity policies were unrelated to the financial crisis. We do not find this  credible, but readers can decide for themselves. We respond to his other points in the detailed response section.

Daniela De Angelis' contribution (with Kandiah and Birrell) makes some points with which we agree, but expresses one opinion with which we do not: that careful model choice, validation and quantification of uncertainty were luxuries that could not be afforded in a crisis. How, in that case, can one avoid the sort of `reckless quantification' that fails the most basic requirements to `first do no harm'? If there is not sufficient validation to know whether a model is remotely suitable for prediction, or whether the reported uncertainties are plausibly well calibrated, then what is a model but a mathematically stated opinion? Or worse, an opinion gaining spurious credibility from the superficial appearance of statistical validity? As in the 2008 financial crisis, we think that managing risk with models whose reliability is oversold has the potential to do great damage, not least because the lure of apparent precision then leads to the down-weighting of other evidence and other factors for which the uncertainties have been more honestly stated.

De Angelis' contribution also makes one substantial technical point that at first sight appears to call into question our results on timing of the incidence peaks (and $R<1$) relative to lockdown. They report results from \cite{kandiah2025} which update \cite{Birrell2021} by allowing contact rates to change weekly before as well as after lockdown, thereby removing the feature of \cite{Birrell2021} that surging incidence up until lockdown was simply built into the model. They report that they still estimate $R$ to be substantially above 1 and incidence to be soaring until lockdown, but they also report results consistent with our paper if they use the same infection to death distribution\footnote{Their figure 3 $R$ results are consistent with ours given the averaging over the week preceding lockdown that their weekly step function entails; their figure 4 is consistent with the regional results in \cite{woodwit2021plos} aggregated in our figure 14. The correspondence would presumably have been clearer if incidence had also been plotted for these scenarios. Note that the northwest is the last region for which $R$ drops below 1.}. This begs the question of what infection to death distribution \cite{Birrell2021} and \cite{kandiah2025} used? Neither paper or the accompanying supplementary material tell us, simply referring to `{\em an assumed-known distribution of the time from infection to death from COVID-19 $f$}' in the Likelihood sections of the supplementary materials. No reference is given, although later in the \cite{Birrell2021} supplementary material it is reported that the mean incubation period was assumed to be 4 days (s.d. 1.41) and the mean time from symptom onset to death was assumed to be 15 days (s.d. 12.1). The papers' cited code repository confirms these numbers and indicates that a gamma distribution was used for the latter distribution, although the  \cite{Birrell2021} code also has an option for a mean 17.8 and s.d. 8.9. \cite{kandiah2025} do not appear to state what the distribution was in the paper or supplementary material, but the \cite{kandiah2025} code uses a gamma distribution with   options of the Birrell parameters (mean 15 with s.d 12.1), or of mean 9.3 with s.d. 9.7, or of mean 9.0 with s.d. 8.1 (this last option is commented `{\em Currently highly unprincipled}'). 

\begin{figure}
\vspace*{-1cm}
\eps{-90}{.5}{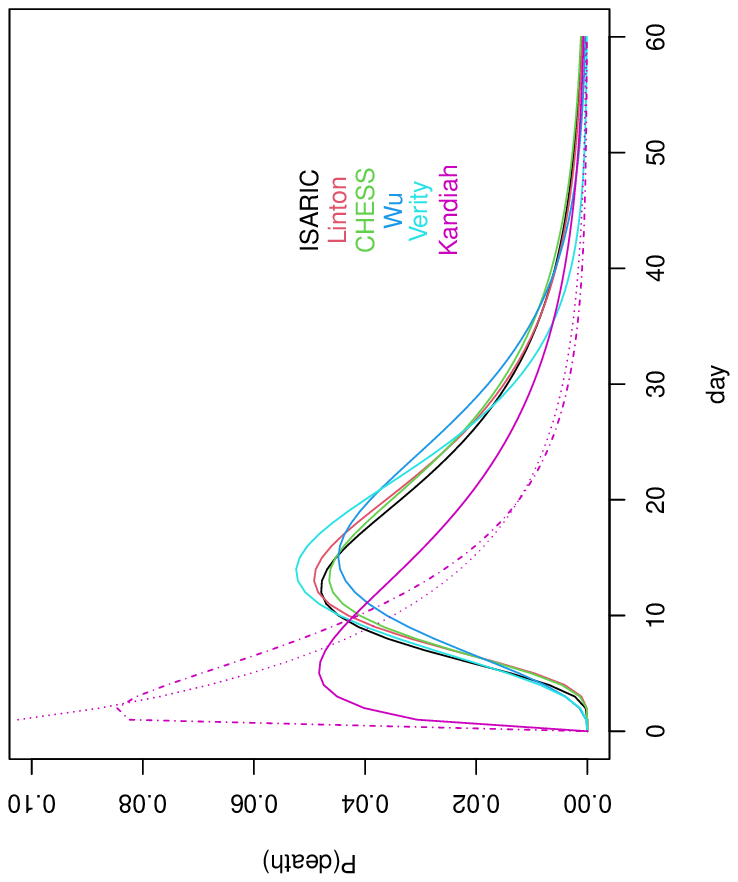}
\vspace*{-.4cm}
\caption{As figure 9 in the main paper, but with the symptom onset to death distributions from the \cite{kandiah2025} code shown in pink. The solid pink curve is also the distribution used by \cite{Birrell2021}. No references or other source information appear to be given for the pink distributions. Note that the unused distribution option in the Birrell code with mean 17.8 would be completely consistent with the plotted distributions from known sources. \label{o2d2-fig}}
\end{figure}

Given the lack of information on where these distributions came from, all we can do is plot them alongside the distributions with known sources given in our paper. Figure \ref{o2d2-fig} shows that they are inconsistent with the distributions from published sources, and the high numbers dying a day or two after symptom onset seem biologically implausible. Additionally, if these distributions were considered reasonable then there is also an awkward mismatch with the incidence estimates based on direct sampling to explain, which does not occur if we use the distributions from published sources. The assumed incubation period distribution also appears shorter than the literature suggests.  

The models used in \cite{kandiah2025} perhaps lean towards a type more common in mathematical biology than statistics. Some important details are parameterized in an ad hoc manner from disparate data sources, with some strong assumptions having to be made. Comparatively few parameters are inferred statistically and it is unclear how that inference is then influenced by the other assumptions and estimates. In particular, key contact processes are {\em assumed} to change at lockdown, apart from those that are estimated. Hence it is unclear to us that the model is really suitable for reliably estimating changes in incidence or $R$ around lockdown, even with more reasonable disease duration distributions. It does not seem suitable for producing the claimed {\em counterfactuals}. Given limitations of space here we would urge the reader to read  \cite{kandiah2025} and \cite{Birrell2021} and form their own view. 

Daniela De Angelis also invited Kevin Fong to present at the meeting. He provided a masterful demonstration of the power of hard hitting emotional messaging over dispassionate consideration of risk, leaving us with the following questions: are there {\em any} limits to the future cost in lives and livelihoods that would be justified to mitigate the trauma to health service staff that a pandemic inflicts? Secondly, does employing emotionally charged messaging to challenge quantitative risk assessments align with the imperative to avoid causing unintended harm?
His written comments are more measured (\url{https://doi.org/10.1093/jrsssa/qnaf109}), underlining that the opening statement of our paper is not hyperbole\footnote{as context to some of his figures: the general mortality rates after ICU admission are around 30\%, with 40\% being about the rate seen for intracerebral hemorrhage ICU patients \citep{van2020long}; health workers suffered Covid death rates around the median of all ONS occupational categories, but this constituted an increase on background death rates by occupation that was (just) at the top of the distribution across the categories.}. What he does not do is to provide any actual evidence that the measures taken were {\em necessary}, nor that there was an appropriate balance between `protect the NHS' and the limitation of other societal damage, including to the development and education of school age children, the future health and wellbeing of the less economically advantaged and the long term economic activity that in the end funds the NHS.  

Many of Sheila Bird's points (\url{https://doi.org/10.1093/jrsssa/qnaf075}) we agree with or are fair comment, but there was also a point about the appropriate multiplier on the NICE threshold and what is appropriate in war-time. We are not sure that the exact multiplier really changes the point that a large discrepancy may suggest something suboptimal. Further, it is the fact that war is societally {\em existential} that justifies the suspension of all usual norms. Presumably no one would still argue that Covid was that, but what if a disease did emerge that was an existential threat? That is both highly contagious and with an infection fatality rate to all ages in the 25-50\% range that is not historically impossible. The NICE threshold would then suggest spending 10-20 times GDP, which is obviously not remotely realistic (it comfortably exceeds total national wealth, for example). The war time analogy raises another awkward question. A contested full scale invasion of the UK would presumably produce casualties at a level that would overwhelm the NHS. Does that mean that the only acceptable option would be immediate capitulation? We agree that of course a pandemic demands exceptional outlay, but that does not remove the need for proportionality. Will it end up being judged proportionate to have taken an economic hit of the order of  \pounds$10^{12}$ combatting Covid, and to have then decided, in the light of the consequent debt burden, that planned spending of around $3\%$ of that, to combat the threat of climate change, was no longer affordable? 

Moving on to some contributions more aligned with the paper's message, Ioannidis, Di Loro et al., Berger et al., and others all discuss the difficulties in getting solid, evidence based narratives heard in the media and elsewhere during the pandemic, with scientific journals also failing to offer the balance and rigour that one might hope.   Ioannidis argues that picking apart how this came about will require a combination of statistics, psychology and sociology. We think he is right. For example, the element of (social) media driven social contagion in the wave of panic over Covid, that swept the developed world and beyond in March 2020, seems at least as significant as the actual risk, serious though that was.  Some of the reason for our susceptibility to such contagion must lie in the obvious adaptive advantages of believing what many other people believe, combined perhaps with a tendency to judge the reliability of others on the basis of their adherence to views we view as sensible. One such view is `the wisdom of crowds’, often supported by Galton’s observation that the median of 800 county fair competitor’s guesses for the weight of an ox was within 1\% of the true weight. But we suspect that an experienced crowd, making an educated guess of a simple quantity, may not be a good model for a social media crowd pronouncing on a complex topic about which  they know little. It is likely true that someone in the crowd will have sensible answers, but equally unlikely that they will get heard above the noise. And that noise was often loudest from the large numbers of suddenly emerging experts, often desperate for the media exposure that might count as academic career enhancing ‘impact’. In this context, a retired GP remarked that he felt that Covid had unleashed an epidemic of the Dunning-Kruger effect.      

Similarly significant is the human desire for control, and the tendency to see events as the result of our actions, even when they are not. One could view the continued insistence on the key role of lockdowns in this light, but in the UK context even the general view of the role of science in the initial lockdown decision perhaps suffers from this illusion of control. At least if Dominic Cummings' evidence to the Covid Inquiry is to be believed \citep{cummings-covid}, the decision to lock down was not primarily driven by the UK government’s official scientific advice (such as Imperial college report 9), but by Cummings and an informal group of four `incredibly able people’ he had turned to. The advisors he lists are indeed outstanding in their respective fields. But those fields do not include epidemiology, public health, statistics or risk management, while, however well read, Cummings' training was in ancient history rather than science.  As Cummings tells it, he was particularly influenced by the advice of a pure mathematician, working with an exponential growth model for Covid, who urged him to  `push extreme suppression immediately’ on 14 March 2020. Cummings then apparently repeatedly warned the prime-minister of the risk of a `zombie apocalypse’, and on 19 March that, without lockdown, `the NHS in London collapses in 15 days’ (c.f. figures 10 or 14 in our paper).
 
The pressures and incentives operating on scientists and decision makers also need serious thought. There is the obvious pressure to make predictions that are broadly correct. The prediction that maximum suppression of human contact will result in suppression of viral transmission is safe in this regard. In contrast any prediction about how little suppression would be sufficient is much harder to get right. The asymmetry in blame is also rather extreme: don’t do enough and you will for sure be seriously blamed. Do too much and it is always possible to argue that it wasn’t. And scientists are human, operating in the same media, social-media and social environments as everyone else. The pressure to do work supportive of what is believed to be right by right thinking people is very high, and, especially in the age of social media, the pressure not to publish contrary work is similarly strong. And that is before we get to the baleful effects of the promotion process on academic freedom of thought.   

To end on a constructive note, and in response to Keeling, Wyse and Srakar's questions about lessons to be learned for the next pandemic, we have some suggestions.
\begin{enumerate}
\item While the Imperial group and others clearly swung into action early, generally the government led scientific response was sluggish, leading to the panic at the heart of government described in Cummings' evidence to the UK Covid inquiry. The obvious lesson for next time is the need to ramp up activity much earlier, so that policy can be based on science, rather than the ad hoc approach described by Cummings.   
\item Do not base quantitative management on un-calibrated models that were not developed for prediction, but rather as very useful aids for structured reasoning and gaining qualitative insight \citep[see e.g.][]{streicher2025plural}.
\item Echoing Britton, Irons and Maruotti et al., risk management should not become focused on a single or small set of objectives simply because competing risks are much harder to quantify with the accuracy claimed for the epidemic itself.
\item Rather than use psychology for manipulating emotions to undermine risk assessment, it would be preferable to focus on the many things that psychology tells us about the confirmation, selection and other biases, to which we are all prey in assessing data and risk, and the way that these can interact with group dynamics to make it difficult to re-assess decisions, once made. In particular, it would probably  be a good idea to have an independent advisory group complementary to SAGE/SPI-M whose {\em remit} is to question the science and to avoid focusing on too narrow or short term a set of societal objectives.   
\item The role of statistics is not to get better parameter estimates to feed into `the models', it is to properly assess (and make decisions given) uncertainty and risk, and provide the methods by which the hypotheses and assumptions encoded in models can be rigorously empirically tested against data. 
\item Statistical methods need to be properly understood to provide useful insight - fitting a model to data is only part of statistical inference, and, as Maruotti et al. emphasize, good fit does not mean good predictive validation. 
 Particularly important are the sampling and design theory necessary for understanding when a set of data form a valid basis for answering a scientific question of interest. Without this, statistical methods can be quite dangerous. For example, likelihood is not a piece of magic to allow models to be fit to data, but fundamentally a model for the random sampling process by which the data are to be treated as being produced.
\item  Measuring a problem is usually better than modelling it: to this end the UK should have a pre-approved surveillance survey and cohort ready to go, with protocols for batch testing thought through for the scenario in which test capacity is limited.
\item It would be good to fill some of the data gaps well before the next pandemic. For example the impressive POLYMOD survey \citep{mossong2008polymod} used to parameterize age specific contact rates in many epidemic models was based on one day diaries kept by around 1000 UK residents of whom 7 were over 75 and none over 80. Such a data deficit should probably be a much higher priority for rectification than method development. 
\item A situation serious enough to justify the closing down of society should surely justify the relaxation of some of the more stringent aspects of data privacy rules to ensure that policy relevant data are open and available for the common good.
\item Try to avoid the yearning for certainty compromising the honest assessment of uncertainty. 
\end{enumerate}
We hope that, despite its necessarily somewhat negative tone, our paper can contribute to the discussion about what the most important lessons were, and that when the next pandemic arrives at least some of  David Spiegelhalter's `many regrets' can be avoided.

\section{Response to individual contributions}

Here we provide brief responses (in alphabetic order of first author last name) to all the discussants, except where the main points are already covered above.  

 \bigskip

\noindent{\bf Allorant} (\url{https://doi.org/10.1093/jrsssa/qnaf099}) argues that we only focused on mortality, ignoring morbidity. Although this is a fair point, we note that we also did not consider morbidity effects of the NPI measures, which are unlikely to have been minor given the tendency for economic deprivation to shorten \emph{healthy} lifespan more than lifespan as such. The argument that mortality maybe underestimated based on  \cite{msemburi2023excess} is extraordinarily weak. See figure 6 of Msemburi, which makes it clear where they believe undercounting occurred and that it is certainly not the UK. Even ignoring all the other evidence for tiny IFR/CFR in the young, what is Allorant's explanation for the low observed deaths in the young if they had high CFR? Were they simply not catching Covid (in which case why lock them down)? Given the unused Nightingale hospitals and the evidence that infections were on the way down before lockdowns we are unsure that the health service collapse was as imminent enough to justify the discounting of future life and wellbeing that took place. And if you fail to adequately prepare the health system for predictable shocks, is it rational to then pay {\em any} price to avoid overloading it -- if not, then where is the threshold?

\bigskip

\noindent{\bf  Barratt.} (\url{https://doi.org/10.1093/jrsssa/qnaf118}) (1) \cite{woodwit2021plos} treat English NHS regions separately. All regions have peak incidence before the first lockdown, but $R>1$ for the North West at lockdown and it is certainly true that London has an earlier peak and earlier $R<1$ point than the rest of the country. Of course this begs the question of why lockdown would be necessary in the rest of the country if the most densely populated and connected region of the UK had infections under control without lockdown? (2) We analyse aggregate dynamics because incidence is additive and the aggregate data have a higher signal to noise ratio. Spatial variability is indeed important: as Diggle points out, a modern surveillance system should be able to provide this. But we believe that the bigger problem remains the imbalance between the attention given to Covid risk versus the long term risks from the response. On (3) we think that the first priority has to be well calibrated communication of risk from the pathogen. If that has been done then locally adapted risk maps are highly desirable. However,  if communication amounts to mapping risk on a scale that runs from `panic' to `panic massively' then it is not so useful.

\bigskip

\noindent{\bf Berger et al.}'s comments (\url{https://doi.org/10.1093/jrsssa/qnaf085}) on the unhelpful sudden emergence of media Covid experts mirror our view of the UK, although the culture war aspects, so well caught in Julia Zeh's novel {\em \"Uber Menschen}, perhaps played out differently in Germany than they did in the UK. Their demonstration of how widespread is the issue of neglected population ageing in excess death calculations is particularly interesting and important. We add this to the lessons learned! 

\bigskip

\noindent{\bf Britton}. (\url{https://doi.org/10.1093/jrsssa/qnaf069}) We agree with the points made, particularly about the necessity of assessing the risk from negative effects of the interventions before and while they are imposed. While we agree that the scientific involvement in measurement and vaccine development in the UK was excellent, we are less convinced that the high level of involvement by many academics was entirely positive, partly because of the dynamic whereby the scientists were focusing only on Covid, and the sheer volume of their work then tended to crowd out the other societal concerns that should have been as or more important. Also, in part, we felt that the UK scientific community developed a sort of herd immunity to data or scientific arguments that challenged the dominant scientific narrative, and this was not helpful.

\bigskip

\noindent{\bf Chadwick et al.} (\url{https://doi.org/10.1093/jrsssa/qnaf116}) We intend no criticism of \cite{hanlon2020} for doing useful work with the data at hand (although their results did lead to the interesting result, at some point in mid 2020, where the average loss of expected life for those who actually died of Covid in the UK was less than the Hanlon estimate, if one only used the victims' age based average expectation of further life!). The injunction to only manage on the basis of facts and data available at the time seems rather difficult in practice: is the fact of co-morbidities being binarized a fact to consider or not? Should the absence of any over 80s in the POLYMOD survey used to parameterize age structured models mean that transmission among the over 80s should be omitted from models, or what?

\bigskip
\noindent{\bf Chind.} (\url{https://doi.org/10.1093/jrsssa/qnaf090}) These comments seem to ignore the fact that if you spend a huge amount helping someone avoid one health risk you can not spend it helping them avoid another, and indeed that if you damage the economy enough in so doing you run the risk of causing long term health and life loss. The NICE threshold is not used at the single patient level. It is about fair allocation of resource at the population level. If one considers the fairness and utilitarian reasons for rationing in this way it is not clear what the basis for completely abandoning these in a pandemic of this nature would be.

We assume that the comments on care workers relate to Peter Goldblatt's contribution. We thought his comments reasonable and not victim shaming.

\bigskip

\noindent{\bf Dagpunar}'s suggestion (\url{https://doi.org/10.1093/jrsssa/qnaf087}) is very interesting but we would caution against giving too much weight to the results. Given the many model simplifications, in the absence of proven good predictive calibration we think it is unwise to give too much credence to extrapolations from these models whether described as predictions, projections or counterfactuals. Leaving aside the important but un-modelled effects of seasonality, the fundamental difficulty is the modelling of complex, variable, ever changing social animals, as if they were molecules of gas. Take Imperial College report 9 - the political difficulty of lifting a lockdown once it was imposed (and people were still dying from Covid) was something completely beyond the scope of the models, yet arguably the critical practical  difficulty with imposing lockdown.

\bigskip

\noindent{\bf De Nicola.} (\url{https://doi.org/10.1093/jrsssa/qnaf102}) We agree completely with these comments and find it reassuring that statisticians working completely independently on this issue arrived at essentially the same conclusions on the importance of accounting for ageing. The fact that their results are so similar based on independently developed and implemented methods differing in detail, further strengthens the importance of the ageing effect. This is independently replicated science.

\bigskip

\noindent{\bf Di Loro et al.} (\url{https://doi.org/10.1093/jrsssa/qnaf081}) offer a cogent analysis of what went wrong with the presentation of risk and science. We were particularly interested in learning about their efforts to counter the misleading narratives disseminated to the
general public during the pandemic. Although statistical literacy is important, if this discussion meeting shows something, it is clear that statistically literate people might come to different conclusions, even with the advantage of hindsight. On the other hand, discussion meetings like these hopefully mean that as a statistical community we can start to work toward a deliberative equilibrium.  

\bigskip

\noindent{\bf Diggle}. (\url{https://doi.org/10.1093/jrsssa/qnaf078}) As we noted in our response to Barratt, we basically agree that having a public health surveillance system that can deliver predictive inferences on key health indicators at fine spatial resolution in near-real time is a good idea.

\bigskip

\noindent{\bf Dodd et al.} (\url{https://doi.org/10.1093/jrsssa/qnaf103}) We note that the credible interval for the Dodd figure 1 is compatible with our results, although we wonder if some artefact of inappropriate assumption of independence has crept in to the \cite{ellison21} method to generate quite such wide intervals. Note also that some of the ageing effect that we account for may not be captured by the Ellison method because of the wide age bands used. We do not think that the suggested cancellation in the ONS figures is what is happening. As we show above, allowing for recent life expectancy trends adds around 28000 excess deaths to our figures (the use of 60 year trends in this context seems unjustified).

\bigskip

\noindent{\bf Dyson} (\url{https://doi.org/10.1093/jrsssa/qnaf072}). Section 1. We agree that robust estimation of the impact of the control measures on Covid is very difficult, but this does not seem to have prevented a widespread very firmly held view that the measures were both effective and proportionate. If the Ferguson et al. projections were good enough in March 2020 to be used as justification for lockdown it is unclear why they would not be good enough for an approximate life year calculation, even with economic forecasts that substantially underestimated the actual costs. Section 2.2: We don't see how the global economic uncertainty can be separated from what individual countries did, especially those that, as Streicher et al. have pointed out, have an outsized reputation and influence in epidemiology. Did the UK government's approach to Omicron have no international impact, for example? Of course the point that different restrictions can have radically different economic impacts is exactly the point.

\bigskip

\noindent{\bf Engebretsen et al.} (\url{https://doi.org/10.1093/jrsssa/qnaf080}) We agree that models fitted to data and used for short term prediction are unlikely to be badly compromised by neglect of individual heterogeneity. But the same can not be said for models from highly influential groups used to predict total death tolls under different policy scenarios and feeding directly into policy. We think that heterogeneity of hospitalization risk is a different issue. While it is true that modelling groups were working very hard, we are less convinced that a seriously flawed model is redeemed by being timely. In this context, it is perhaps worth mentioning that in the interval between the acceptance and publication of \cite{flaxman2020lockdown}, Nature rejected \cite{wood2021peaktiming}, without any question as to its technical correctness.

\bigskip

\noindent{\bf Fang.} (\url{https://doi.org/10.1093/jrsssa/qnaf089}) It seems difficult to argue that the reductions in personal contact prior to lockdown in the UK and Sweden were caused by lockdown, given that Sweden did not lockdown, even if one is unconcerned by the effect preceding the cause. Unfortunately our Norwegian is not good enough to comment meaningfully on the Statistics Norway report \citep[][only the one page summary is in English]{statsnorway2022}.  
However the point remains that Sweden is an example of how the pandemic could be handled without full lockdowns and the full economic and social damage that these entailed. A comparison of the change in Swedish and UK government debt levels after Covid at least suggests that the ongoing costs and opportunity costs would have been lower under a less stringent approach.      

On cases and waste water, we refer the reader to Figure 2 of \cite{fang2022wastewater}. 

\bigskip

\noindent{\bf Fanshawe}. (\url{https://doi.org/10.1093/jrsssa/qnaf077}) Certainly MD in Private Eye was wise before the event and we don't think he was alone. The point about risk is that it includes uncertainty. If you pile up poor quality studies you tend to exaggerate in one direction. Honest risk communication would not be stressing rather speculative dangers, but would, we think have been closer to the MD column from before lockdown. Point out that there is a risk of post viral complications, but also that virtually everyone was going to get Covid eventually and it was highly uncertain that any vaccine would protect against post viral complications.

Fanshawe is right that we got it wrong about blinding in the Matta study (an example of multiple rounds of revision not always leading to improvement).  We did not cover \cite{stephenson2023long} as we were not aware of it when first writing this section in early 2023. While another example of a properly conducted prospective study it is affected by the same drop out issues we highlight (as the authors acknowledge), and it is perhaps worth quoting the paper's conclusions {\em in full}.
\begin{quote}
In CYP [Children and Young People], the prevalence of specific symptoms reported at time of PCR-testing declined with time. Similar patterns were observed among test-positives and test-negatives and new symptoms were reported six months post test for both groups suggesting that symptoms are unlikely to exclusively be a specific consequence of SARS-COV-2 infection. Many CYP experienced unwanted symptoms that warrant investigation and potential intervention.
\end{quote}
We note that Fanshawe does not comment on the ONS long term impact figures - are these too an example of us downplaying the risk and if so how?

\bigskip

\noindent{\bf Fisch} (\url{https://doi.org/10.1093/jrsssa/qnaf070}). On point 1 see the discussion from Di Loro et al. Point 2 is covered in the main response. On point 3 we are not clear how the ONS {\sf Reff} is obtained before the ONS survey started, and we could not find the relevant information on the link provided. On point 4 we note that during the apparently harsh Swiss measures the ski resorts and borders were open. 

\bigskip

\noindent{\bf Fisher} (\url{https://doi.org/10.1093/jrsssa/qnaf113}) We completely agree with this, and especially the need for measurement to happen as soon as it becomes apparent that there might be a problem. We tried to make the measurement point early on in 2020, but were utterly naive about how to go about it, or how much noise would have to be made to have a hope of being heard.

\bigskip
\noindent{\bf Goldblatt}'s, clearly heartfelt, contribution (\url{https://doi.org/10.1093/jrsssa/qnaf111}) mirrors the objections of the referee whose views did most, over 4 refereeing rounds and a year and a half, to ensure that our paper would emerge too late to contribute to the national debate surrounding the Covid enquiry's consideration of the science. Taking the points in turn. Accounting for population ageing and the baby boomers is not somehow separate from the life table approach: the latter is what needs to be done to achieve the former. Fundamental to point 1 is a belief that austerity was unrelated to the financial crisis --- we do not believe this is credible, but readers can decide for themselves. On point 2, there is obviously little point in computing excess deaths to assess the effects of the epidemic if you have {\em defined} Covid deaths as excess a priori. In addition since dying of any cause means that you die earlier than you would otherwise have done then Goldblatt's definition gets close to defining all deaths as excess. Point 3 seeks to disagree with the point we make about the effects of harvesting within a 3 year period and to support this then makes exactly the same argument for a 100 year period: it is difficult to respond to logic of this sort. On point 4 we obviously did not argue that lockdowns were unnecessary because cases and deaths are lagged data etc. With verbal models of the sort then presented you can prove anything. On point 5, to argue that self reporting exaggerates rates is not to argue that the underlying rates are zero: we do not negate. Point 6 presumably does not require rebuttal for anyone who has actually read section 3 of our paper. 

\bigskip

\noindent{\bf Gomes et al.}'s (\url{https://doi.org/10.1093/jrsssa/qnaf104}) demonstration of the potential to leverage regional differences in initial conditions to address the identifiability issues that arise when modelling heterogeneity is particularly appealing. Combined with better measurement of individual heterogeneity in contact rates, we think that this work points the way to adequately dealing with the heterogeneity issue in future (exactly the sort of methods development that Keeling calls for).  

\bigskip

\noindent{\bf Gupta}'s (\url{https://doi.org/10.1093/jrsssa/qnaf101}) point, that when you combine the seasonality, heterogeneity and behavioural changes and acknowledge the underestimation of exposure inherent in serology data (a point also made by Fanshawe), the turn around in infection rates before lockdown is unsurprising and may have as much to do with biology as behaviour is important. The desire for control is human, but it repeatedly pushes us to ascribe to our own actions what nature and chance have given us little real control over.    

\bigskip

\noindent{{\bf Hall}}  (\url{https://doi.org/10.1093/jrsssa/qnaf076}) In response to the points we could follow: 1. Does this mean that risk distortion is justified? 3. Seems to conflate case data and random samples. See what REACT2 actually did. 4. See the time dilation approach in Wood (2021).

\bigskip

\noindent{{\bf Held et al.} (\url{https://doi.org/10.1093/jrsssa/qnaf106}) gives a nice overview of the rather rigorous statistical approach that partly underpinned what we view as the less damaging Swiss measures. While Switzerland is of course also a rich country with a well resourced health system, we also wonder if the generally high standard of technical literacy within government may also have played a part in achieving somewhat better risk management. Purely anecdotally, in August 2021 EW spoke at the Swiss Statistics Days conference in the same session as Alain Berset, Swiss President of the Council of Ministers. With Covid still ongoing, Berset's team were very keen to get the preprint of \cite{woodwit2021plos}, in some contrast to the work's UK reception. In contrast, during Covid the UK science and technology committee of the house of commons had 3 members with a science background and Boris Johnson's then chief advisor, Dominic Cummings, has a degree in ancient history.

\bigskip

\noindent{\bf Hill} (\url{https://doi.org/10.1093/jrsssa/qnaf071}) disagrees somewhat on the utility of case data. Of course this depends on what you are doing, but we believe there was far too much emphasis given to these data of convenience once actual measurements became available.  

\bigskip

\noindent{\bf House} (\url{https://doi.org/10.1093/jrsssa/qnaf115}). That Medley commented on Gomes on Twitter seems weak evidence of the heterogeneity effect being taken seriously. House's comments elide heterogeneity of susceptibility/transmissibility with household or network structure. This is odd given that \cite{house2011cluster} take pains to avoid such confounding, by setting the degree of each node in their network models to be constant. The nice paper by \cite{pellis2020structure} focuses on populations structured into children and adults and household structure, and not person-to-person variability (see their Methods section). The \cite{dattner2021covid-kids} data is on transmission within households during lockdown: its relevance to the question of heterogeneity in normal contact rates is unclear. So the cited references do not appear to suggest that the heterogeneity we discuss does not have a unidirectional effect, especially so in the context of the simple SEIR type models that were actually calibrated against data statistically.

\bigskip

\noindent{\bf Ioannidis.} (\url{https://doi.org/10.1093/jrsssa/qnaf079}) In all honesty we can not disagree with the points made. A public health emergency is not the moment at which to abandon the evidence based approach, but much of the response, from the untested use of lockdowns to the absurd restrictions on outdoor activity, did just that.

\bigskip

\noindent{\bf Irons} (\url{https://doi.org/10.1093/jrsssa/qnaf083}) is entirely correct that the alternative to the measures taken was never going to be business as usual: there would be a substantial economic hit in any case. But the point of drawing attention to the figures as we do is to emphasise that the consequences were so large that the `spend whatever it takes' attitude would be bound to cause immense future hardship. In the UK that is evident now in the massive extra post pandemic debt servicing payments that inevitably can not be spent on health, energy transition, education or any number of other socially desirable areas.  

\bigskip

\noindent{\bf  Johnson.}  (\url{https://doi.org/10.1093/jrsssa/qnaf094}) We don't really understand how distinguishing targetted and mass testing relates to the distinction between cases and prevalence. Even if one agrees that the right hand panel of Johnson's figure 2 would be an acceptable error in a derivative in this context, we don't see how the number of tests administered each week relates to the assumption that those tested are somehow a random sample of those tested, or the concern that the ascertainment fraction is likely not to be constant. The problem with high volume Twitter commentary is that it is rather easy to pick out where it turned out to be right, after the event. If we recall rightly, {\tt @BristOliver} was less prescient when it came to hospital load from Omicron.

We agree with Johnson that lockdown almost certainly prevented additional Covid deaths during the first wave (whether the price was then paid in the second wave, as modelling suggested it would be, is another matter). But basing the cost of lockdown on the cost of furlough alone seems rather economically naive, and takes no account of the economically mediated future life loss, which, while hard to predict, appears from previous data to be a substantial risk.

\bigskip

\noindent{\bf Jose}'s comments (\url{https://doi.org/10.1093/jrsssa/qnaf086}) on risk communication are important, which is why we think that David Spiegelhalter's original explanation of risk in terms of background risk in a year was so well done, and why we concur with him that it is a pity that it did not form the basis for the approach to risk communication taken. Whatever one's level of numeracy, one is likely to have some level of understanding of how one feels about risk of dying this year and hence to gain some useful understanding. A school friend of SNW's daughter commented in autumn 2020 ``the way it was presented I should have known loads of people who died from it and I don't know anyone" -- that this feeling was anything but unique, suggests the sort of risk communication failure that can only erode trust in science in the future. We agree too with the comments on models. What did the UK do well? Once it got going, the ONS survey that actually measured the state of the epidemic was excellent, and we remain baffled by the failure of other countries to replicate this. REACT and REACT2 were also superb and the openness of much data was also exemplary. The central UK government not pursuing the pseudoscience of zero Covid and, perhaps through growing model skepticism, declining to lock down for Omicron were also huge positives. The UK's lab based biomedical scientists, like those of several other countries, obviously did a great job on vaccines.

\bigskip

\noindent{\bf Keeling} (\url{https://doi.org/10.1093/jrsssa/qnaf073})  complains that we chose to pick a comparison in which models did badly, and there were many cases where they did well. The difficulty here is that the available tests are rather asymmetric: it is easy to predict that if you suppress hard then cases will come down, and much harder to predict either the outcome under reduced measures, or how little suppression would have sufficed. But management by model requires that the model is capable of making all these predictions, not just the heavy suppression ones. We can all agree that a sledgehammer will crack the nut, but how small a hammer could we have used? The predictions for Sweden and Omicron are the real tests that we have of the ability to get close to this latter question. On hindsight and Omicron, SNW was interviewed for Scottish television at the time, pointing out that the Omicron models being used appeared to be highly pessimistic chimera, but getting heard at this point at the time was difficult (as was contributing to policy at all, unless one bought in to the dominant narrative).

Keeling asks for the lessons to be learned for the next pandemic, and we have listed some in the main points section. Keeling himself emphasizes the need for new method development, but we think that such developments are perhaps less important than the need for new measurements (and perhaps better communication of what existing methods can and can not do). For example the POLYMOD survey \citep{mossong2008polymod} used to parameterize age specific contact rates in many epidemic models was based on one day diaries kept by around 1000 UK residents of whom 7 were over 75 and none over 80. That is a data deficit that should probably be a much higher priority for rectification than method development.   

\bigskip

\noindent{\bf Kumar} (\url{https://doi.org/10.1093/jrsssa/qnaf084}) makes an important point about dark data, of which perhaps the most troubling early example was the WHOs presentation of case fatality rates without the appropriate level of emphasis on the dark data of undetected non-fatal cases. Another is the missing data on social contact rates between those in the most vulnerable age groups. It is possible that lockdowns had some short term life prolonging effects, of which the most obvious would be the interruption of the normal circulation of respiratory pathogens that are often among the causes of death for the very elderly. Against this must be stood the reduction in access to healthcare for those suffering from life threatening conditions such as heart attack, sepsis and stroke - the conditions that usually keep ICUs near capacity. At least in 2020 ONS data on deaths at home were consistent with rates for these conditions having increased. 
  
 \bigskip

\noindent{\bf Liu et al.}  (\url{https://doi.org/10.1093/jrsssa/qnaf100}) We think that the convolution process that smooths incidence trajectories as they become sytmptom onset, hospitalisation and death trajectories calls for considerable caution in modelling using step functions aligned with policy announcements. At very least all conclusions made using them need to be checked against alternatives that assume more continuous dynamics. Similar caution is needed when assuming smooth trajectories of course, and \cite{wood2021peaktiming} spent considerable effort on checking the effects of assuming a smooth trajectory. The same does not seem to have been true for the analyses assuming step changes. With the hindsight provided by the REACT2 and ONS incidence measurements it is at best unclear that step functions or similar are appropriate for incidence.

\bigskip

\noindent{\bf Maruotti et al.} (\url{https://doi.org/10.1093/jrsssa/qnaf096}) We agree completely with the need for serious consideration of the very real trade-offs and the crucial importance of the difference between good model fit and good predictive validation \citep[as emphasised in the context of epidemic models in][for example]{wood1999super}.

\bigskip

\noindent{\bf Molenberghs.} (\url{https://doi.org/10.1093/jrsssa/qnaf093}) The \cite{abrams2021covid} model cited as an example of models including person to person variability is simply a stochastic implementation of a model without person-to-person variability except that related to age (section 2.2 of Abrams gives the model), perhaps emphasising the poor understanding of this issue. Belgian debt has not risen as dramatically as in the UK, so the economic effects may well be less (debt servicing costs are another matter, of course). Earlier intervention is obviously better if all you care about is the immediate problem. But earlier interventions will presumably need to be in place for longer and create a larger problem the following respiratory virus season. Hence it is entirely unclear that they reduce, rather than increase the economic damage and consequent health and societal damage. Particularly unclear is that this is rational risk management before it is clear that a vaccine is possible. 

\bigskip

\noindent{\bf Pellis} (\url{https://doi.org/10.1093/jrsssa/qnaf117}).  Because of the profoundly damaging and untested nature of the intervention we would not have imposed lockdown 1, viewing the downside risks as too high and the intervention as having no viable exit strategy given the uncertainty over the possibility of a vaccine; this was especially so given the obvious political difficulty there would be in lifting a lockdown in a timely manner. However we think this decision was finely balanced and other views were reasonable, given the uncertainties and weak health service capacity. However what was profoundly wrong was the duration of the lockdown, which was irrational unless the scientific fantasy of zero Covid was being pursued.

The deconvolution analyses were first produced in April 2020, in plenty of time to have been used in the management of the last pandemic, let alone the next, but then, as now, there was a reluctance to take the implications on board. 

The deconvolution method is not a GAM. One can do an approximate version of it using GAMs, but this involves assuming smoothness on the raw incidence scale, which is obviously suboptimal. The fact that a method is mechanism agnostic does not mean that it can not detect a pattern generated by a mechanism, nor diminish the utility of detecting it as a model sanity check (notice that the MRC method also detects the pattern).

We are not sure what useful purpose is served by Pellis' figures 1 and 2. The smooth in figure 1 is obviously the wrong model for the data presented, so the $R$ estimates resulting from it will obviously not be correct.   

We completely agree that fine control of fast-growing systems with long delays is really hard. Doing it on the basis of mathematical models never designed or validated for the task is even harder. That is why we think that in the next pandemic early measurement should be the biggest priority: the testing capacity was there this time.

\bigskip

\noindent{\bf Rougier}. (\url{https://doi.org/10.1093/jrsssa/qnaf126}) The first lockdown is where the distribution of time to death is likely to be most reliable, given that the distribution was estimated from fatal hospitalized cases up to October 2020. The exception to this is that we used the distribution from time to infection to death for the general population and there is evidence that this period is longer in the older patients making up most of the deaths \citep{tan2020age-inc}. This may partly explain why our peak is a little later than the REACT2 based estimates. Later on we think that the point is of more concern: by the time selective vaccination is changing the demographic profile of the victims there is obviously a worry, even without variant changes. But by then there are direct sources of information for the UK, which suggest that the approach is at least not estimating peak incidence to be earlier than occurred in the general population.

The second suggestion is very elegant, but we think that the all excess deaths are from Covid assumption is difficult. Paraphrasing a medic from a large ICU, to the first author in 2020: `it was tough, we were full with Covid patients. But the thing is those beds are not normally empty, they are full of patients with sepsis, heart attacks and stroke. I've talked to colleagues all over the country and it's the same. The question is where have those other patients gone? The assumption is they are dead'. The ONS excess deaths at home data at the time did not contradict this.            

\bigskip

\noindent{\bf Selby}. (\url{https://doi.org/10.1093/jrsssa/qnaf095}) On $R$, \cite{wood2021peaktiming} includes more sources of uncertainty, but we agree that outside London it is unclear if $R$ was below 1 before lockdown (or even if discussing the instantaneous value of a quantity playing out over future time makes sense, at least with regard to relative timing differences well below the generation time). Perhaps there is the sort of backwards in time causality from lockdown suggested, perhaps people had a last minute burst of activity before being locked down, or perhaps, like the authors, most people really didn't know what the prime-minister would actually announce until he did so. Behaviour that changes $R$ instantly, changes incidence instantly. We're not sure what was wrong with the qualitative prediction of the models that higher first wave suppression would lead to larger second waves.

\bigskip

\noindent{\bf  Senn} (\url{https://doi.org/10.1093/jrsssa/qnaf068}) is right - when we went back to the data on which the Marmot report plot is based it does appear that there is an error, as he points out (although it does not change the conclusions). We should have gone back to the original data before. This is in fact the only point at which we used `second hand' data. Other data used were either from data files provided by the stated source, or sometimes digitized from the data producers own publications.

\bigskip

\noindent{\bf Singh}. (\url{https://doi.org/10.1093/jrsssa/qnaf082}) On long Covid we think it prudent to at least adjust the quoted prevalences using the results from the studies looking at the proportion of self reported long Covid against seroprevalence. We agree that the discussion of economic effects is difficult, uncertain and crude, but none of those things are reasons to discount the effects in favour only of the effects presented as being more quantitatively accessible.  

\bigskip

\noindent{\bf Spiegelhalter} (\url{https://doi.org/10.1093/jrsssa/qnaf105}). We agree with these comments and especially the need for intervals to reflect realistic calibrated uncertainty. If defensible levels of uncertainty accompanied epidemic model predictions then there would perhaps be less tendency to down-weight the highly uncertain downside risks when balancing policy options. See also the response to Jose.

\bigskip

\noindent{\bf Srakar} (\url{https://doi.org/10.1093/jrsssa/qnaf119}) asks what we need in order to be better prepared in future. Infection surveys ready to go for a pre-prepared cohort. Much stronger training in the fundamentals of what makes the statistical approach valid, so that we do not become blinded by methods into forgetting the data.

\bigskip

\noindent{\bf Streicher and Broadbent} (\url{https://doi.org/10.1093/jrsssa/qnaf088}) have been modest in promoting their own work, but of the many papers cited by discussants, \cite{streicher2025plural} is one of the ones that most repays reading. Particularly salient is their cogent argument that $R<1 $ is simply the wrong target in practice. Their other work highlighting, for example, the inequities of the pandemic responses outside the first world is also worth seeking out. It is also relevant to the UK, given, as they argue, the disproportionate role of UK groups and institutions in framing and setting the global response. For example, when Neil Ferguson, of Imperial College's highly influential WHO collaborating Department of Infectious Disease Epidemiology, realized that it was possible to `get away with'\footnote{In a December 2020 interview with Tom Whipple in the Times, Ferguson is quoted as saying, with reference to lockdown in Wuhan: {\em ``It’s a communist one-party state'' we said. ``We couldn’t get away with it in Europe,'' we thought. And then Italy did it. And we realised we could.} } lockdown in the UK, it was likely to have had worldwide effects well beyond the UK. 
\bigskip

\noindent{\bf Tian et al.}  (\url{https://doi.org/10.1093/jrsssa/qnaf107}) The wide versus narrow caliber issue is indeed a serious one. In the UK context death certification guidelines were changed early in the first wave to include Covid on death certificates if it was suspected that it could be involved: as one medic put it in the context of care home deaths: `now everyone dies of Covid, previously we didn't speculate on the specific pathogen if someone dies of pneumonia, now it's all Covid'. Carl Heneghan successfully campaigned to get the the UK government to move from an absurdly wide definition that would have counted a road death as Covid if the victim had ever had Covid, but the variability even in the definition of Covid death is still a problem. In response to point (2), we are not sure why the $t$ distribution is deemed incorrect here. Code is provided for fitting the models in the supplementary material. On (3), see the contribution from Gomes and associated papers.

\bigskip

\noindent{\bf Tildesley} (\url{https://doi.org/10.1093/jrsssa/qnaf074}). It is slightly surprising that those who express most scepticism about the possibility of predicting the no-lockdown counterfactual are the most convinced that lockdown was necessary. The verbal model that the vulnerable will peak before the rest of the population is difficult to reconcile with the ONS or REACT2 results which suggest that the fatal incidence peak was after the general incidence peak. The latter phenomenon is perhaps expected. We used the infection to symptom distribution for the general population, whereas there is evidence \citep{tan2020age-inc} that older people actually have significantly longer incubation periods: this effect would tend to shift the inferred fatal peak to later than it should be. Additionally, at least for SEIR type dynamics, sub-populations with lower contact rates tend to peak later than those with higher rates, not earlier, for reasons related to those discussed in Section 4.1 of our paper.   

\bigskip

\noindent{\bf Vencalek} (\url{https://doi.org/10.1093/jrsssa/qnaf098}) provides a very nice case study  illustrating  how a pandemic does not in anyway remove the considerable difficulties that attend observational data, and the need, if anything, for extra care in the pandemic situation.

\bigskip

\noindent{\bf Whitehouse et al.} (\url{https://doi.org/10.1093/jrsssa/qnaf097}) are surely right to identify problems of identifiability  as a central difficulty in dealing with the substantial effects of heterogeneity on disease models, although as Gomes shows elsewhere in the discussion, the difficulty can be overstated. But whatever the level of practical difficulty, the question is how to fill the information gap in future? A start would be to devote substantial effort to measurement of the heterogeneity in contact rates. The UK models relied heavily on the POLYMOD survey: a stupendous multi-country effort to look at contact rates, but for the UK still based on around 1000 one day contact diaries, with 7 participants over 75 and none over 80 \citep{mossong2008polymod}. Given the risk profile of Covid and the seriousness of the situation, this is clearly inadequate (patching a nuclear reactor with ductape?). A much larger statistical study is needed. Even then the question of how to adapt the measured baseline rates as behaviour changes in the face of risk remains as a challenge. 

\bigskip

\noindent{\bf  Wyse et al}. (\url{https://doi.org/10.1093/jrsssa/qnaf110}) Obviously the deconvolution approach is retrospective, but the analysis from section 5 was first applied at the end of April 2020. The results about peak timing were the same as now. The paper was rejected without review by JRSSC in early May 2020 for (perhaps disingeneously) `insufficient methodological novelty'. For the incubation period the first version used \cite{lauer2020covid}, which was online 10 March. MacAloon's paper was on medRxiv in April 2020. So while it is true that the analysis is retrospective, it was in practice available quite early enough to have contributed to some adjustment of the dominant narrative and to policy, had there been willingness to take its message on board: instead there was a tendency for the results to be dismissed as a `right wing talking point'\footnote{the Google scholar record for the preprint changed from pointing to the arXiv to the `Daily Stormer' a couple of weeks after posting (slightly surprising given that Google had apparently stopped indexing the `Daily Stormer' some years previously, or perhaps not, given what is written in Matt Hancock's diaries).}. See also figure 2 of \cite{wood25pt2} for an investigation of how soon reliable inference could feed policy. See the main response for some of the other improvements we suggest. On sampling, the ONS and REACT surveys show how to do it, but it would be a good idea to have pre-approved studies ready to go at the first hint of pandemic possibility.

\end{document}